\documentclass[a4paper,fleqn]{cas-dc}

\usepackage[numbers]{natbib}

\usepackage{layout}
\usepackage{subfig}
\usepackage{xspace}
\usepackage{tabularx,makecell}
\usepackage{multirow}                 
\usepackage{multicol}                 
\usepackage[section]{placeins}        
\usepackage{float}
\usepackage{amssymb}
\usepackage{amsmath}
\usepackage{cleveref}
\usepackage{url} 
\usepackage{hyperref} 
\usepackage{orcidlink}

\newcommand{\methodALL}{{\textit{FlowGen}}\xspace}
\newcommand{\methodBFGen}{{\textit{BFGen}}\xspace}
\newcommand{\methodBPPredictor}{{\textit{BPP}}\xspace}
\newcommand{\methodAFGen}{{\textit{AFGen}}\xspace}

\crefformat{figure}{#2Fig.~#1#3}
\crefformat{table}{#2Table~#1#3}  
\crefformat{section}{#2Section~#1#3}  
\crefformat{equation}{#2Eq.~(#1)#3}  

\def\tsc#1{\csdef{#1}{\textsc{\lowercase{#1}}\xspace}}
\tsc{WGM}
\tsc{QE}

\begin{document}

\ExplSyntaxOn
\cs_set:Npn \__first_footerline:
{
  \group_begin:
  \small
  \sffamily
  \ifnum\theblind>0\relax
  \else
    \__short_authors:
  \fi
  \group_end:
}
\ExplSyntaxOff
\let\WriteBookmarks\relax
\def\floatpagepagefraction{1}
\def\textpagefraction{.001}

\shorttitle{Relationally Guided Use Case Modeling with LLMs}

\shortauthors{Guangyu Wang et al.}

\title [mode = title]{Relationally Guided Use Case Modeling with LLMs}

\author[1,2]{Guangyu Wang\orcidlink{0009-0001-8130-9135}}
\ead{wgy@buaa.edu.cn}

\author[1]{Bangqi Li\orcidlink{0000-0001-6591-3204}}
\ead{libangqi@buaa.edu.cn}

\author[1,3]{Ji Wu\orcidlink{0000-0002-3937-2368}}
\cormark[1] 
\ead{wuji@buaa.edu.cn}

\author[1]{Zhijun Shao\orcidlink{0000-0002-6154-4664}}
\ead{zjshao@buaa.edu.cn}

\affiliation[1]{organization={School of Computer Science and Engineering},
            addressline={Beihang University},
            city={Beijing},
            postcode={100191},
            state={Beijing},
            country={China}}

\affiliation[2]{organization={Xi'an Aeronautics Computing Technique Research Institute},
            addressline={AVIC},
            city={Xi'an},
            postcode={710065},
            state={Shaanxi},
            country={China}}

\affiliation[3]{organization={Engineering Research Center of Integration and Application of Digital Learning Technology},
            addressline={Ministry of Education},
            country={China}}

\cortext[1]{Corresponding author}

\begin{abstract}
Use case flows are important elements of use case modeling because they support downstream software engineering activities, including requirements analysis, architectural and detailed design, and test case generation. 
However, constructing them manually is costly and expertise-intensive, while existing automated approaches still struggle to preserve semantic consistency, control-flow logic, data-flow logic, and the intended system boundary, especially when identifying branch points and generating alternative flows. 
To address this problem, we propose \methodALL for complete use case flow construction. 
\methodALL uses LLM-based Semantic Information Processing (SIP) to extract semantic elements, constructs a Semantic Relational Graph (SRG) encoded by an enhanced R-GAT for basic flow generation (\methodBFGen), and further supports branch point prediction through \methodBPPredictor and branch-conditioned alternative flow generation through \methodAFGen.
Evaluations on 13 public and 7 industrial datasets show that \methodALL consistently outperforms competitive baselines in all three core components. In particular, \methodBFGen improves over the best baseline by 14\% in Precision, 7--25\% in Recall, 11--30\% in F1, and 10--19\% in AUC; \methodBPPredictor improves Precision by 30--110\%, Recall by 33--91\%, and F1 by 32--117\%; \methodAFGen improves Precision by 8--23\%, F1 by 5--18\%, and AUC by 0.6--2.5\%.
Moreover, we validate the effectiveness of the LLM-based SIP module and the attention preservation factor in \methodBFGen, analyze the impact of requirement completeness on \methodBFGen, and examine how different scopes of branch-related context affect \methodAFGen.

\end{abstract}


\date{}

\maketitle

\section{Introduction}
Use case modeling is widely used to capture and organize high-level functional requirements as structured interaction scenarios between actors and a target system \cite{wiegers2013software,tiwari2015systematic}.
Such scenarios support downstream analysis, software design, and testing activities, such as requirements analysis \cite{yue2013facilitating}, architecture and detailed design \cite{vranic2024usecase,liu2025ucd,santos2025requirements}, and test case generation \cite{wang2020automatic}. 
In this paper, we focus on automating the construction of use case flows, which constitute the central behavioral part of a use case specification, including the basic flow that describes the main success scenario and the alternative flows that handle exceptional or deviating situations. 
In particular, we consider branch points, namely the steps at which a base flow, i.e., the basic flow or another alternative flow, should branch into a corresponding alternative flow.

Although engineers can author such flows manually, doing so is time-consuming and depends heavily on domain expertise \cite{lian2025reqcompletion,ko2019automatic,wang2024test}. 
Over the past decades, a series of automated methods have been proposed to support use case flow construction \cite{elrakaiby2022care,jahan2021generating,yue2015atoucan,jurkiewicz2015automated,20163402722828,al2018use,8663869}. These methods have improved automation to a considerable extent, but their limitations have become more evident as software requirements grow more complex and more domain-specific.
In particular, many earlier rule-based approaches rely heavily on handcrafted templates, syntactic patterns, and sentence-level parsing, and are therefore effective only in restricted domains or input styles. More importantly, they often fail to capture cross-sentence semantics, branch-triggering conditions, and inter-step data-flow dependencies. For alternative flow generation, many methods treat the base flow as plain text and derive divergent scenarios through predefined patterns \cite{20163402722828,makino2012scenario}, which weakens the explicit connection between a branch point and the corresponding exception-handling flow.

Recent advances in deep learning \cite{sanyal2025hybrid,al2018use}, especially Large Language Models (LLMs) \cite{chaaben2023towards,krishna2024using}, have opened new possibilities for automating use case flow construction. Compared with rule-based methods, LLMs provide stronger natural language processing capabilities and can better exploit domain knowledge through pretraining \cite{jin2024large,10.1145/3695988}. Nevertheless, LLM-based generation is still inadequate for this task. 
First, LLMs may fail to fully capture the contextual dependencies needed for constructing structured software artifacts from requirements, including use case flows, especially when the input is long or structurally complex \cite{jin2024evaluation,liu2024llms,lal2024catbench}.
Second, they may generate plausible but out-of-scope behaviors, thereby violating the system boundary of the target use case. 
Third, they remain weak at branch-point reasoning. In particular, deciding whether a given step should trigger an exceptional response, and maintaining semantic and logical continuity between that triggering step and the generated alternative flow, requires conditional and causal reasoning that current LLMs often fail to perform reliably \cite{liu2024llms,chi2024unveiling}. As a result, generated alternative flows may be misaligned with the base flow from which they originate.

These limitations indicate that automating use case flows is not merely a text generation problem. 
The core challenge is to preserve the logic encoded in the requirement.
Preserving this logic is essential because use case flows guide subsequent requirements analysis, software design, and testing; errors in the generated flows may therefore propagate to these downstream activities.
This logic has at least four aspects. 
First, the generated flow should preserve \emph{semantic consistency}: actions should remain aligned with the relevant entities, data items, and domain terms described in the requirement. 
Second, it should preserve \emph{control-flow logic}: the flow should follow the intended sequencing and diverge at the appropriate branch points. 
Third, it should preserve \emph{data-flow logic}: the availability and state of data across steps should constrain what later steps can legitimately do.
Fourth, it should preserve the \emph{system boundary}: generated behavior should remain within the intended scope of the target use case rather than drifting to related but out-of-scope functionality.

To address these challenges, we propose \methodALL, which comprises three modules: Basic Flow Generation (\methodBFGen), Branch Point Prediction (\methodBPPredictor), and Alternative Flow Generation (\methodAFGen). 
\methodBFGen combines semantic extraction through LLM-based Semantic Information Processing (SIP), Semantic Relational Graph (SRG) construction, and enhanced R-GAT encoding to make semantic elements, flow dependencies, and branch-triggered relations explicit. \methodBPPredictor models branch-triggering control-flow logic in base flows, and \methodAFGen uses branch-local context together with globally encoded use case semantics to maintain continuity between branch points and generated alternative flows, thereby helping the generated flows remain within the intended system boundary. 
To support branch-point prediction, we also introduce a semi-automatic pipeline that leverages LLM suggestions and expert validation to annotate branch points in public datasets.

We evaluate \methodALL on 13 public datasets and 7 industrial datasets. 
Overall, \methodALL consistently outperforms competitive baselines across basic flow generation, branch point prediction, and alternative flow generation, with average gains over the best baselines of 38.60\% in Precision, 40.07\% in Recall for the tasks where Recall improves, 38.92\% in F1 score, and 8.16\% in AUC. 
The comparisons against LLM-based baselines and the Sequence Transformer baseline indicate that SRG-based graph conditioning is useful for preserving branch-to-flow alignment and requirement logic.
The results further confirm the effectiveness of the SIP module and the attention preservation factor in \methodBFGen, show that \methodBFGen remains robust under incomplete requirements, and demonstrate that using an appropriate branch-context scope contributes to \methodAFGen.

The main contributions of this paper are as follows:
\begin{itemize}
\item We extend our previous \methodBFGen \cite{wang2025bfgen}, which addresses only basic flow generation, to \methodALL for complete use case flow generation by further introducing branch point prediction and alternative flow generation;
\item We introduce \methodBPPredictor for identifying branch points in base flows by explicitly modeling branch-triggering control-flow logic;
\item We design \methodAFGen, an alternative-flow decoder that integrates branch-local context with globally encoded use case semantics to produce flows coherent with their triggering branches while remaining within the intended system boundary;
\item We design a semi-automatic annotation pipeline to extend publicly available datasets from 8 application domains by supplementing branch point annotations to enable \methodBPPredictor.
\end{itemize}

The paper is structured as follows: \cref{sec:related work} summarizes and positions our work relative to prior studies. \cref{sec: approach} presents the framework of our proposed approach \methodALL and its details. \cref{section:experiments} describes our experimental setup, including datasets, evaluation metrics, research questions, and experimental setting. \cref{sec: results} presents our experimental results and analysis. \cref{sec: Threats to Validity} discusses potential internal, external, and construct threats to our study. \cref{sec: conclusion} concludes the paper and outlines directions for future work.

\section{Related Work} \label{sec:related work}
\subsection{Use Case Flow Construction}

Automatically deriving use case flows from high-level natural-language requirements has long been studied as an important problem in use case modeling.
Among existing approaches, rule-based methods combined with NLP techniques are some of the most frequently cited solutions.
These approaches rely on syntactic patterns and rules, such as extracting action-object pairs through part-of-speech tagging, to parse requirement sentences, identify primary actions and associated entities, and subsequently construct event flows using predefined templates  \cite{al2018use,jahan2021generating,wang2020automatic,yue2015atoucan,alashqar2021automatic}.
To address the deficiency in action and entity extraction caused by natural language ambiguity, researchers have encoded domain knowledge into predefined rules \cite{8663869,wang2025assessing,jin2024evaluation}.
However, such rules typically lack the necessary generalizability across domains. In contrast, \methodALL leverages LLM-based semantic extraction to identify domain-specific concepts, actions, and entities in a data-driven manner, achieving better cross-domain generalizability.

Several studies have incorporated formal methods, machine learning, or neural models to improve automation \cite{elrakaiby2022care,al2018use,ko2019automatic}. For example, Elrakaiby et al. \cite{elrakaiby2022care} refine requirements through a calculus-based process guided by refinement operators. Al-Hroob et al. \cite{al2018use} employ NLP tools and neural networks to extract actors and actions from requirements. Ko et al. \cite{ko2019automatic} use verb clustering and external knowledge to detect omitted steps in use case scenarios. 
Although these methods improve automation for specific subtasks, several of them still require non-trivial manual effort, such as defining refinement operators, validating extracted elements, or preparing external knowledge sources. Most of them also focus on local syntactic evidence or task-specific heuristics. In contrast, our work aims at a more automated pipeline that preserves the logic of use case construction beyond local syntactic evidence, including sequential control-flow logic and data-flow dependencies among actions and objects.

Recently, there has been increasing interest in applying LLMs to requirements engineering \cite{10.1145/3695988}. Existing studies mainly confirm the LLM's potential of requirement specification generation \cite{krishna2024using}, and explore tasks such as requirement elicitation \cite{ren2024combining}, requirement specification mining \cite{20230146707}, and high-level requirements modeling \cite{chaaben2023towards}. These studies show that LLMs are useful for understanding requirement text, but they also expose persistent limitations such as weak domain grounding and hallucination \cite{wang2025assessing,jin2024evaluation,li2023survey,jin2024large}. 
More importantly, existing LLM-based studies have not directly addressed structured use case flow construction, where flow continuity and action-object consistency must be preserved explicitly \cite{liu2024llms,jin2024evaluation}.

Beyond generating the main success scenario, complete use case flow construction also requires identifying where the base flow may diverge into exception-handling behavior.
To the best of our knowledge, few studies have specifically addressed the task of branch point prediction to derive alternative flows. Some research efforts on event identification in use cases are implicitly relevant to the recognition of branching conditions \cite{jurkiewicz2015automated,williams2022automated,rago2016assisting,makino2008method}.
Jurkiewicz and Nawrocki \cite{jurkiewicz2015automated} proposed an approach for automatically identifying potential exceptional events in the main scenario of a use case. Their method extracts actors, activities, and data objects from the action steps in the basic flow and predicts exceptional events using inference rules inductively derived from existing use cases. Their method heavily relies on manually predefined data object attributes, and its effectiveness is constrained by both the predefined inference rules and the limited coverage of the existing use cases.
Williams et al. \cite{williams2022automated} developed an ontology-driven framework that recommends potential security concerns by inferring likely security threats through ontology reasoning and predefined concern mappings of the actors, actions, and assets, and their relations specified in use case specifications.
However, the effectiveness of inferring security threats relies on the manual efforts: annotating ontology instances in use case specifications, defining a security concern taxonomy, and validating the inferred threats.
In contrast, \methodALL predicts branch points by capturing semantic information and structural dependency from the use case description and the base flow context, without requiring manually constructed domain-specific knowledge or ontology instance annotations.

Most approaches to alternative flow generation rely on rule-based methods applied to use case descriptions \cite{shudo2010method,makino2012scenario,20163402722828}.
Ko et al. \cite{20163402722828} proposed a pattern-based method to systematically refine use case descriptions by suggesting possible alternative scenarios.
To automatically recommend potential branches or exception paths, this method introduces a catalog of use case specification patterns to capture the typical relations between the basic flow and alternative flows.
However, its effectiveness is restricted by the manually defined specification patterns. 
In addition, the provided patterns focus primarily on structural or syntactic information rather than semantics in use case specifications.
Unlike pattern-based methods, \methodALL combines LLM-based semantic extraction with explicit branch point prediction instead of relying on predefined templates. The predicted branch context is then used to guide alternative flow generation, thereby better preserving the semantic and logical continuity between a base flow and its corresponding alternative flow.

\subsection{Graph Neural Networks in Software Engineering}

Graph neural networks (GNNs) have been widely adopted in software engineering because they are effective at modeling structured relations that are difficult to capture with purely sequential text encoders. Prior studies have used GNNs for software modeling \cite{10602548}, model recommendation \cite{di2021gnn,10589780}, effort estimation \cite{20224212974810}, clone detection \cite{mehrotra2023improving}, and fault localization \cite{rafi2024towards}. The common advantage of these methods is that they exploit graph structure to integrate local and global dependencies into learned representations.

Our work builds on this observation but uses GNNs in a different way. Rather than modeling code structure or issue relations, we model requirement-to-flow logic. 
The SRG used in \methodALL represents use case descriptions and flows as a task-specific typed and weighted graph over extracted semantic elements, with relations capturing intra-flow dependencies, description-to-flow grounding, and branch-triggered cross-flow connections. This representation is designed specifically for use case flow generation, where semantic consistency, flow continuity, and exception-handling logic are all central.

The effectiveness of GNNs in software engineering tasks depends on the scale and quality of labeled datasets \cite{20214711206403}. However, such datasets are scarce in practice, particularly for domain-specific scenarios, due to the high cost of manual annotation. To address this challenge and enable \methodALL, we designed a semi-automatic pipeline to annotate and complete multiple public datasets.

\section{Our Approach: \methodALL} \label{sec: approach}

\begin{figure*}[h]
\begin{center}
\includegraphics[width=0.9\linewidth]{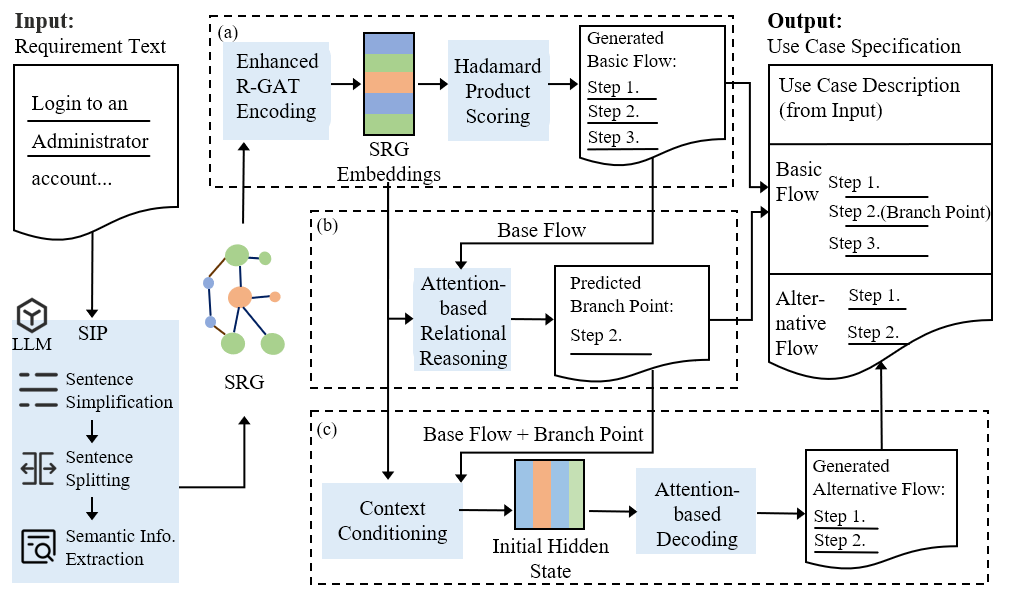}
\end{center}
\setlength{\abovecaptionskip}{0pt} 
\caption{The Overall Process of \methodALL. SIP: Semantic Information Processing; SRG: Semantic Relational Graph; (a): Basic Flow Generation; (b): Branch Point Prediction; (c): Alternative Flow Generation.}
\label{pic: The Overall Process of}
\vspace{0cm} 
\end{figure*}

In this section, we introduce the pipeline of \methodALL, as illustrated in \cref{pic: The Overall Process of}.
In \cref{sec:Construction}, we formally define the elements extracted from requirement descriptions and use case flows, and present the method for constructing an SRG, laying the foundation for \methodALL. 
In \cref{sec:sip}, an LLM-equipped Semantic Information Processing module extracts the defined elements and constructs the SRG. 
This graph is then fed into the \methodBFGen module (\cref{section:bfgen}) to generate the basic flow.
Building upon the SRG and the base flow, potential branch points are predicted (\cref{sec:bppredictor}) and used to guide \methodAFGen in generating alternative flows (\cref{sec:afgen}).

\subsection{Semantic Relational Graph Construction} \label{sec:Construction}

\begin{figure*}[h]
\begin{center}
\includegraphics[width=0.8\linewidth]{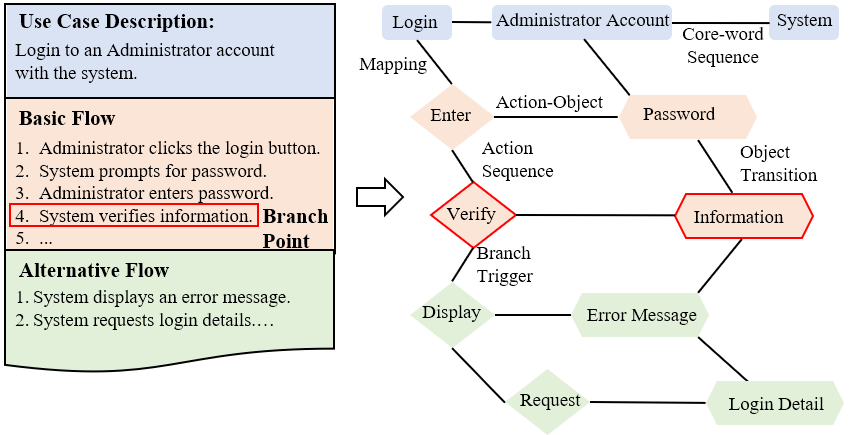}
\end{center}
\setlength{\abovecaptionskip}{0.cm} 
\caption{Illustrative Semantic Relational Graph Fragment. Blue nodes denote core words, orange nodes denote basic flow actions/objects, green nodes denote alternative flow actions/objects, and red borders indicate the branch point. Edges illustrate representative relations, including action/object/core word sequence or transition ($E_1/E_2/E_6$), action-object relation ($E_3$), description-to-flow mapping ($E_4/E_5$), and branch-triggered cross-flow relations ($E_7/E_8$).}
\label{pic: Illustrating An Example of Graph}
\vspace{-0.5cm} 
\end{figure*}

In this subsection, we define the semantic elements extracted from a use case and the relations used to construct the SRG. In this work, we adopt the SRG as a task-specific typed and weighted dependency graph to expose the requirement logic required by the downstream modules.

As illustrated in \cref{pic: Illustrating An Example of Graph}, the functional requirement description of a use case is usually presented in natural language, where domain-specific terms, actions, and content words are used to describe what the use case performs.
Formally, the textual content of a use case description ($Desc$) can be represented as $n$-tuples:
\begin{equation}
  Desc = < c_{1}, c_{2}, \ldots, c_{n} >, c_{i} \in C
  \label{eq:Desc}
\end{equation}
where \textbf{$C$} denotes the set of core words appearing in the use case description, to specify what the use case does. 
A use case specification ($UCS$) consists of one basic flow $BF$ and zero or more alternative flows $AF$. Each alternative flow stems from one branch point $BP$.
\begin{equation}
    UCS=BF \cup \bigcup_{k=1}^{n} AF^{(BP_k)}
\end{equation}
The basic flow $BF$ of a use case is presented as a sequence of action steps. Each step typically consists of an action and its associated objects, specifying the stepwise behavior. 
\begin{equation}
    BF = [<a_{1}, o_{1}>,...,<a_{p}, o_{p}>],\ a_{i} \in A, o_{i} \in O, 1 \leq i \leq p 
    \label{eq:bf}
\end{equation}
where $p$ is the total number of steps in the basic flow, \textbf{$A$} is the set of actions, and \textbf{$O$} is the set of objects involved in actions.
A branch point $BP$ refers to a specific action step within a $BF$ or an $AF$, where system status or input might have exception(s) that need to be handled in the corresponding $AF$. 
Taking the branch points in the basic flow as an example: $BP=\{BP_1,...,BP_k \}, 0 \le k\le p$.
Each alternative flow $AF^{(BP_k)}_i$ consists of $m_k$ action steps and specifies how the exception triggered at $BP_k$ is handled:
\begin{equation}
\begin{split}
    AF^{(BP_k)}_i = [<a_{1}, o_{1}>,...,<a_{m_k}, o_{m_k}>],
    \\a_{j}\in A, o_{j}\in O,1 \leq j \leq m_k 
\end{split}
    \label{eq:af}
\end{equation}
In the graph representation, each step, whether it is a step in the basic flow, a step in the alternative flow, or a branch point, is operationalized through its constituent action and object nodes. 

We group the relations in the SRG into three categories:

\textbf{Intra-flow dependency relations.}
${E_1 \subseteq A \times A}$: Sequential relation of two consecutive actions in a basic flow or an alternative flow; ${E_2 \subseteq O \times O}$: Transitional relation of two data objects accessible in a common data flow; ${E_3 \subseteq A \times O}$: The action-object relation between an action and an object, meaning the action accesses the object and applies to the action steps in both basic flow and alternative flows.
These relations capture the local logic of control flow and data dependency, together with the action-object semantic consistency within a flow.

\textbf{Description-to-flow grounding relations.}
${E_4 \subseteq C \times A}$ \& ${E_5 \subseteq C \times O}$: The mapping relation between a core word in the functional description of a use case and an action or object in the corresponding basic flow and alternative flows; ${E_6 \subseteq C \times C}$: Sequential relation of two core words in the functional description of a use case. $E_4$, $E_5$, and $E_6$ connect the requirement description to the flow representation by linking core words to actions and objects, and by preserving the sequential context among core words in the description. These relations ground generated flows in the semantics of the original requirement and help preserve the system boundary of the use case.

\textbf{Branch-triggered cross-flow relations.}

\noindent${E_7 \subseteq A_{BSF} \times A_{AF}}$: The triggering relation between the branch point and the first action of the corresponding alternative flow.
${E_8 \subseteq O_{BSF} \times O_{AF}}$: The triggering relation between the object in the base flow and the object in the corresponding alternative flow.
$E_7$ and $E_8$ are introduced to explicitly model the logic of control flow and data dependency between the triggering step and the exception-handling flow derived from it.

In practical datasets, some action sequences, object transitions, and cross-flow correspondences occur repeatedly within a domain and reflect stable domain conventions.
For example, in cloud service applications, the "activate service" action typically follows the "deactivate service" action, and both frequently appear in related use cases.
Similarly, in applications with security concerns, such as banking or e-commerce, the "authentication" action almost invariably triggers the "grant access" action. Both actions consistently work on fixed objects---such as user credentials or session tokens---and these objects appear more frequently than others in these applications.
We therefore associate each relation with a frequency-based weight. The weight does not claim that the relation is universally more correct; instead, it serves as a salience prior indicating how prominent that relation is in the observed requirements and flows of the domain \cite{sparck1972statistical,manning1999foundations}. In the definition $w(e)=k \cdot f(e)$, the constant $k$ controls scaling, while the relative importance among relations is determined by their observed frequencies $f(e)$.

Based on the extracted nodes, typed relations, and relation weights, the use case can be specified as an SRG $G = (V,E,W)$, where $V = C \cup A \cup O$ denotes the set of core word, action, and object nodes, $E$ is the set of typed edges constructed from the relations among these nodes, and $W$ contains the corresponding weights.
This graph serves as the structured representation on which the subsequent modules perform logic-aware generation and prediction.

\subsection{Semantic Information Processing} \label{sec:sip}
In this subsection, we present the Semantic Information Processing (SIP) module, designed to extract the defined semantic information from requirement descriptions and use case flows. SIP has three sequential tasks: Sentence Simplification, Sentence Splitting, and Semantic Information Extraction.

Software requirements in natural language often include modifiers and compound sentences to describe the scenarios or behaviors in which the software system interacts with its actors. This significantly increases the difficulty of extracting semantic information from requirements using NLP tools or LLMs \cite{narayan2014hybrid}.
To improve the effectiveness of semantic information extraction, \methodALL first simplifies sentences and then splits compound sentences into simple ones.
The Sentence Simplification task removes unnecessary modifiers from a sentence while preserving its core words. The Sentence Splitting task splits a compound sentence into simple ones without changing the sequence of verbs and objects.
Subsequently, \methodALL extracts semantic information: (1) core words from the use case description, such as domain terms, actions, and other content words; (2) actions and objects from action steps in use case flows.

To improve the accuracy and efficiency of extraction, we use LLMs with strong natural language processing capabilities \cite{key,baiduERNIE40Turbo128KModelBuilder}.
Building upon existing structured prompt frameworks \cite{jin2024evaluation,krishna2024using}, we modularize the prompt into three components, forming a triplet $\mathcal{P} = (\mathcal{R}, \mathcal{T}, \mathcal{I})$:

\textbf{Role Description} ($\mathcal{R}$): This component assigns a role to the LLM, guiding it to apply task-specific knowledge during processing to align with the task domain.

\textbf{Task Description} ($\mathcal{T}$): This component provides a comprehensive task description and output schema to ensure the results can be processed consistently in downstream tasks.

\textbf{Inputs} ($\mathcal{I}$): This component specifies the concrete inputs of the task. The input of the Sentence Simplification task is a requirement description or use case flow, which is then fed to the Sentence Splitting task. The Semantic Information Extraction task takes the result of the splitting task as input.

\begin{figure}[]
    \centering
    \includegraphics[width=1.0\linewidth]{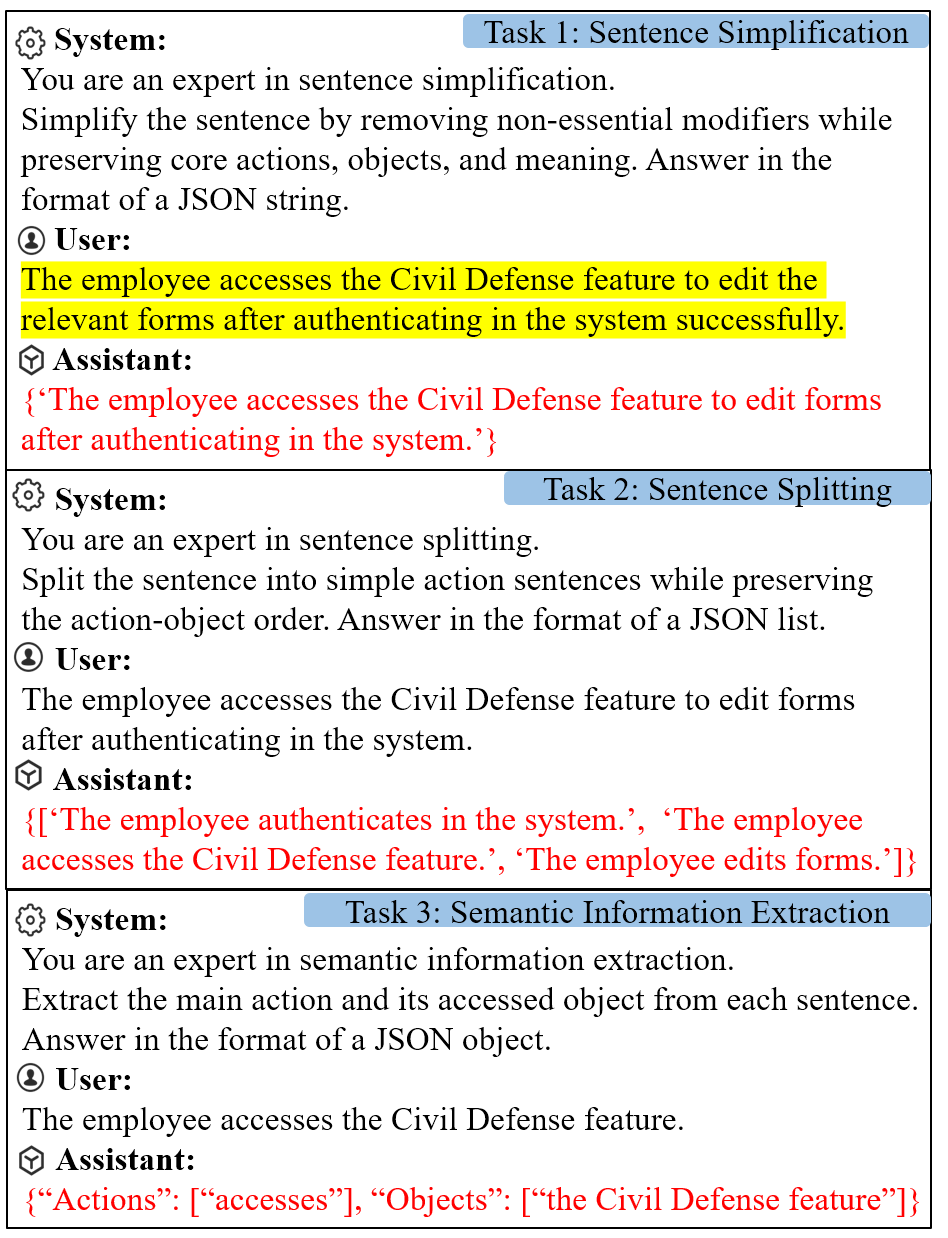}
    \setlength{\abovecaptionskip}{0.cm} 
    \caption{An Example of The Prompt of Semantic Information Processing}
    \label{Fig. prompt}
    \vspace*{-0.5cm} 
\end{figure}

\cref{Fig. prompt} demonstrates the execution of tasks using an action step statement from a basic flow in the eANCI \cite{hey2021improving} dataset, where the original sentence is highlighted in yellow and the LLM outputs in red.
In Task 1, the LLM removes adverbs such as "successfully" in the input as they are deemed non-essential to the core meaning of the step. 
In Task 2, the LLM splits the compound sentence into three simple ones based on the action execution order derived from semantic analysis. 
In Task 3, the simple sentences from Task 2 are processed to extract actions and objects, resulting in (action, object) pairs such as “\textbf{accesses}” (action) and “\textbf{the Civil Defense feature}” (object).
Another scenario in Task 3 involves extracting core words from use case descriptions. This requires adding a domain-specific term extraction instruction to Task Description $\mathcal{T}$ while keeping other components unchanged.
As shown in \cref{pic: Illustrating An Example of Graph}, the term "\textbf{Administrator account}" is identified alongside other core words (e.g., "\textbf{Login}") via the extraction task.

With the results of the Semantic Information Extraction task, \methodALL establishes relations among the extracted semantic elements and assigns weights based on the relation definition in \cref{sec:Construction}. Specifically, $E_1$ and $E_2$ capture the sequences of actions and objects respectively within the action steps of the use case flow;
$E_3$ captures actions and the accessed objects within action steps; 
$E_4$ and $E_5$ capture the mapping between the use case description and the corresponding use case flow, linking each core word to each action and object; 
$E_6$ captures the sequence of core words within the use case descriptions;
$E_7$ and $E_8$ capture the mapping relations between the base flow and the alternative flows at branch points.
Relation weights follow the frequency-based salience prior, i.e., $w(e)=k\cdot f(e)$ with $k=1$ by default.

\subsection{Basic Flow Generation (\methodBFGen)} \label{section:bfgen}
After constructing the SRG and extracting semantic information through the SIP module, \methodALL generates the basic flow using the \methodBFGen module, which serves as the structural backbone for the entire use case flows. \methodBFGen is designed to capture the contextual dependencies among actions and objects, and to predict a coherent sequence of action steps that specifies the main scenario of a use case.
\methodBFGen leverages an enhanced Relational Graph Attention Network \cite{meng2023rgat} to encode the SRG. By integrating neighborhood information, relation types and weights into the attention mechanism through a multi-layer structure, \methodBFGen enables each node embedding to not only represent its local semantics but also capture long-range dependencies of different types and weights within a use case context.
We introduce an attention preservation factor that explicitly regulates the balance between attention-based embedding aggregation and the uniform embedding aggregation. This factor is incorporated into the message-passing mechanism of R-GAT, yielding the following enhanced propagation rule:
\begin{equation}
h_i^{l+1} = \sigma\left( \sum_{j \in \mathcal{N}(i)} \left( \lambda \alpha_{ij} + (1 - \lambda) \right) W^l h_j^l \right)
\label{eq:node_update}
\end{equation}
where \( h_i^{l+1} \) is the updated embedding of node \( i \) at layer \( l+1 \), obtained by aggregating the transformed embeddings from its neighboring nodes in \( \mathcal{N}(i) \) at the \(l \)-th layer. 
\( W^l \) is a trainable parameter matrix, and $\sigma(\cdot)$ is a nonlinear activation function.
\( \lambda \in [0,1] \) is the attention preservation factor. By dynamically adjusting the impact of $\lambda$, the model can flexibly capture important relations while preventing overfitting to noise or weak associations.
The attention coefficient \( \alpha_{ij} \) for the edge from node \(j \) to \( i \) is computed as shown in \cref{eq:attention_coeff}:

\begin{equation}
\alpha_{ij} = \frac{\exp\left( \sigma\left(r_{ij} \mathbf{a}^\top \left[ W^l h_i^l  \|  W^l h_j^l \right] \right) \right)}{\sum_{k \in \mathcal{N}(i)} \exp\left( \sigma\left(r_{ik} \mathbf{a}^\top \left[ W^l h_i^l  \|  W^l h_k^l \right] \right) \right)} + w_{ij}
\label{eq:attention_coeff}
\end{equation}
where $r_{ij}$ and $r_{ik}$ denote the edge representation associated with edges $(i,j)$ and $(i,k)$, \( \mathbf{a} \) is a learnable parameter vector, and \( w_{ij} \) denotes the edge weight.

\methodBFGen employs scoring functions, as shown in \cref{eq:score_cal} and \cref{eq:score_avg}, combined with binary cross entropy (BCE) loss \cite{hurtik2022binary} to predict nodes and train the model. For each input, i.e., use case description, \methodBFGen computes the probability across all action/object nodes as the prediction relevance score and outputs the set of positive nodes whose scores exceed the predefined threshold $\theta$.  
Specifically, let \( C_{Desc} \) denote the associated core word nodes that occur in a functional description \( Desc \), and let \( V = A \cup O \) denote the union of action and object nodes from basic flows in the training set. For each core word node \( c \in C_{Desc} \) and each action/object node \( v \in V \), given that R-GAT has a total of $L$ layers, the Hadamard product of the final-layer representations, \( \mathbf{h}_c^{L} \) and \( \mathbf{h}_v^{L} \), is used with sigmoid activation to generate a relevance score:  
\begin{equation}
s_{c,v} = \text{sigmoid} (\mathbf{h}_{c}^{L} \odot \mathbf{h}_{v}^{L})
\label{eq:score_cal}
\end{equation}
The final score $s_v^{(Desc)}$ for each action/object node \( v \) of \( Desc \) in the basic flow is obtained by averaging over all \( s_{c,v} \).
\begin{equation}
s_v^{(Desc)} = \frac{1}{|C_{Desc}|} \sum_{c \in C_{Desc}} s_{c,v}.  
\label{eq:score_avg}
\end{equation}
The embeddings $H_{BF} = [h_1^{BF},h_2^{BF},...,h_p^{BF}]$ represent the semantics of the most contextually and domain-aligned set of actions and objects, collectively forming a sequence of action steps that constitute the corresponding basic flow for a given use case description.

\subsection{Branch Point Prediction (\methodBPPredictor)} \label{sec:bppredictor}
Following the generation of the basic flow, \methodALL proceeds to generate alternative flows to handle potential exceptions. As a crucial prerequisite, it first identifies potential branch points, i.e., specific action steps within a use case flow where exceptions may trigger conditional or exceptional alternative flows. Although a branch point is defined conceptually at the action-step level, each step is represented in the SRG through its constituent action and object nodes. Accordingly, \methodBPPredictor performs node-level learning.
We design an encoder–decoder architecture, where the encoder leverages the enhanced R-GAT introduced in \cref{section:bfgen} to generate contextual embeddings, and the decoder, implemented as the \methodBPPredictor module, assesses the likelihood of each step being a branch point.
Specifically, \methodBPPredictor adopts an attention-based relational reasoning mechanism, where each flow node (action/object) attends to neighboring nodes in the graph to infer its importance.
Given the encoded node embeddings $H = [h_1,h_2,...,h_n]$ produced by the encoder, and the embeddings of the base flow nodes $H_{BSF} = [h_1^{BSF},h_2^{BSF},...,h_n^{BSF}]$, the probability of each node embedding is computed as:
\begin{equation}
    P(BP_i)=\sigma\!\left(W_p[h_i^{BSF}\Vert\text{Attn}(h_i^{BSF},H)]+b_p\right)
    \label{eq:P(BP_i)}
\end{equation}
where $\text{Attn}(h_i^{BSF},H)$ denotes the attention-weighted contextual embedding aggregated from all nodes in $H$, $\sigma$ is the sigmoid activation; $W_p$ and $b_p$ are trainable parameters. The model is trained with BCE loss:
\begin{equation}
    \mathcal{L}_{BP}=-\frac{1}{p}\sum_{i=1}^{p}[y_i\log P(BP_i)+(1-y_i)\log(1-P(BP_i))],
\label{eq:loss bp}
\end{equation}
where $y_i \in \{0,1\}$ indicates whether the node $v_i^{BSF}$ is a branch point. $v_i^{BSF}$ is recognized as a branch point when $P(BP_i) \geq \tau$, where $\tau$ is an empirical threshold. 
The predicted branch points $BP=\{BP_t\}$ and their contextual embeddings are then provided as input to the Alternative Flow Generation (\methodAFGen) module (\cref{sec:afgen}), which constructs the sequence of action steps to handle the potential exceptions (as responses) branching from the identified branch points.

\subsection{Alternative Flow Generation (\methodAFGen)} \label{sec:afgen}
Based on the predicted branch points, the base flow in which they are located, and the use case description, \methodAFGen generates alternative flows using an encoder–decoder architecture.
Consistent with \methodBPPredictor, the encoder reuses the contextual embeddings produced by the enhanced R-GAT in \cref{section:bfgen}, ensuring all input elements reside in a shared semantic space.
The decoder, distinct from that of \methodBPPredictor, comprises two key components: 
(1) Context Conditioning, which fuses the context around the branch point with the semantics of the first node in the alternative flow into a unified initial representation, so as to anchor how the divergence starts from the identified branch.
(2) Attention-based Decoding, which incrementally and iteratively constructs the step sequence of the alternative flow by attending to the initial representation, along with the local context and global use case semantics.

\textbf{Context Conditioning}
The decoder performs autoregressive generation, where each action step in the alternative flow is generated sequentially based on the previously generated steps and the initial hidden state.
The initial hidden state of the decoder $h_{z_{0}}$ encodes contextual information and initializes the autoregressive decoding process by combining two components: $\hat{h}_{C_i}$ that captures the local semantics around the branch point, and the first node (action/object) embedding in the alternative flow, denoted $h_{AF_0}$. 
During training, the first node of the reference alternative flow is used as a start-node signal to reduce the semantic gap between the identified branch point and the beginning of the target alternative flow.
$\hat{h}_{C_{i}}$ is computed by aggregating node representations in the local neighborhood of the branch point. 
Specifically, for a given branch point $BP_i$, contextual information is derived from its $k$-hop neighboring flow nodes, where the parameter $k$ controls the scope of the neighborhood.
\begin{equation}
    \hat{h}_{C_{i}}=\frac{1}{\left|\mathcal{N}_{k}(BP_i)\right|} \sum_{v \in \mathcal{N}_{k}(B P_i)} h_{v}
\end{equation}
where $\mathcal{N}_{k}(BP_i)$ denotes the set of nodes reachable from $BP_i$ within $k$ hops (including $BP_i$ itself), and $h_{v}$ is the embedding of node $v$ produced by the enhanced R-GAT encoder.

$h_{z_{0}}$ is computed by concatenating $\hat{h}_{C_{i}}$ with the first node of the alternative flow $h_{AF_0}$ at step $t = 0$, allowing the decoder to maintain access to the context encoded in the enhanced R-GAT while inheriting the local semantics of the branch.
\begin{equation}
    h_{z_{0}}=\operatorname{ReLU}\left(W_{h}\left[\hat{h}_{C_{i}} \Vert  h_{AF_0}\right]\right),
\end{equation}
where $W_h$ is a learnable transformation matrix and $[||]$ denotes vector concatenation.

\textbf{Attention-based Decoding}
At each decoding step $t$, the decoder generates the next node, attending to the entire context encoded by the enhanced R-GAT.
The attention weight $\alpha_t$ over encoder representations $H = [h_1,...,h_n]$ is computed as:
\begin{equation}
\begin{aligned}
\alpha_{t} &=\operatorname{softmax}\left(w^{\top} \tanh \left(W_{a}\left[h_{z_{t-1}} \Vert H\right]\right)\right), \\
c_{t} & =\sum_{j=1}^{n} \alpha_{t}^{(j)} h_{j}, \quad h_{z_{t}}=\operatorname{GRU}\left(\left[x_{t-1} \Vert c_{t}\right], h_{z_{t-1}}\right),
\end{aligned}
\end{equation}
where $w$ is a trainable parameter vector used to compute attention weights over the encoder states, $x_{t-1}$ is the embedding of the previous output token and $c_t$ is the attention-weighted context vector.
Starting from the initial state $h_{z_0}$, the decoder recurrently updates its hidden state $h_{z_t}$ as it processes each generated step in the alternative flow sequence.

The output probability distribution for the next token $y_t$ is computed by projecting the hidden state $h_{z_t}$ onto the candidate set of actions/objects:
\begin{equation}
    p(y_t \mid y_{<t}, AF_i) = \text{softmax}(W_o h_{z_t})
\end{equation}
where $W_o$ is a learnable weight matrix that maps the context-aware hidden state $h_{z_t}$ to logits over the output candidate set. 
The softmax function normalizes the resulting scores to yield a valid probability distribution over potential next actions or objects, conditioned on the generation history $y_{<t}$ and the specific alternative flow $AF_i$ context.
\methodAFGen is trained to minimize the negative log-likelihood of the target sequence, formulated as:
\begin{equation}
\mathcal{L}_{\text{decoder}} = 
- \sum_{t=1}^{T} \log p(y_t^* \mid y_{<t}, AF_i),
\label{eq:loss af}
\end{equation}
where $y_t^*$ denotes the reference token at step $t$.
The encoder-decoder architecture and the above computational mechanism enable \methodAFGen to integrate local branching cues and long-range dependencies within the SRG, ensuring that the generated alternative flows are accurate and semantically consistent with both the requirement and the corresponding base flow.

\section{Experiments} \label{section:experiments}
In this section, we detail the experimental design and setup used to evaluate the effectiveness of \methodALL. We first introduce the collected public and industrial datasets, and the branch point annotation in \cref{subsection: Datasets}. Subsequently, we introduce the specific evaluation metrics for generation and prediction tasks in \cref{Evaluation Metrics}. Finally, we outline the seven research questions (RQs) that guide our investigation in \cref{sec:rq set}, and specify the experimental environment and parameter settings in \cref{sec: experiment environment}.

\subsection{Datasets} \label{subsection: Datasets}

\begin{table*}[h]
    \centering
\setlength{\abovecaptionskip}{0.cm} 
    \caption{Overview of Public and Industrial Datasets\\Note: BF = Basic Flow; AF = Alternative Flow; Sys. = System; O\&M = Operations \& Maintenance.}
    \scalebox{0.9}{  
    \begin{tabular}{l l c c c c c c c}
    \hline
    &\textbf{Dataset Name} & \textbf{Domain} & \textbf{\makecell{\# of\\Use Cases}} & \textbf{\makecell{\# of\\BF Steps}} & \textbf{\makecell{\# of BF\\Unique Steps}} & \textbf{\makecell{\# of AF}} & \textbf{\makecell{\# of\\AF Steps}} & \textbf{\makecell{\# of AF\\Unique Steps}} \\
    \hline
    \textbf{\multirow{20}{*}{\makecell{Public\\Datasets}}} &  
    \textbf{EasyClinic} & \makecell[c]{Laboratory\\Management} & 30 & 210 & 187 &58 & 200 & 113 \\ \cline{2-9}
    &\textbf{hats} &\makecell[c]{GUI for Program\\Transformation} & 28 & 183 & 162& 45&	163&	117\\\cline{2-9}
    &\textbf{eANCI} & Governance & 139 & 602 & 480 & 0 & 0 & 0\\\cline{2-9}
    &\textbf{keepass} &\makecell[c]{Password\\Management} & 14 & 66 & 56 & 38	& 81 & 61\\\cline{2-9}
    &\textbf{eTour} & Tourism & 58 & 272 & 230 &0&0&0\\\cline{2-9}
    &\textbf{viper} & \makecell[c]{Supply Chain\\Management} & 39 & 105 & 92 &18	&23	&18\\\cline{2-9}
    &\textbf{iTrust} & Healthcare & 34 & 356 & 346 &43	&57	&50\\\cline{2-9}
    &\textbf{gamma j} & Web Store & 26 & 132 & 104 &14	& 44	& 25\\\cline{2-9}
    &\textbf{inventory} &\makecell[c]{Inventory\\Management Sys.} & 10 & 35 & 33 &12	&34	&24\\\cline{2-9}
    &\textbf{inventory 2.0} & \makecell[c]{Inventory\\Management Sys.} &21 & 191 & 184&21	&48	&35 \\\cline{2-9}
    &\textbf{SMOS} & Education & 67 & 214 & 197 & 0 & 0 & 0 \\\cline{2-9}
    &\textbf{model manager} & \makecell[c]{Research Task\\Management} & 7 & 81 & 81 &0&0&0\\\cline{2-9}
    &\textbf{pnnl} & Diagnostic & 5 & 45 & 45 &9	&34	&33\\
    \hline
    \textbf{\multirow{13}{*}{\makecell{Industrial\\Datasets}}} &
    \textbf{01} & Sys. Platform &111 & 1142 & 408 &319	&638	&155\\\cline{2-9}
    &\textbf{02} & \makecell[c]{Disaster Recovery\\ \& Backup} &5 & 58 & 15 &35	&70	&8 \\\cline{2-9}
    &\textbf{03} & \makecell[c]{Network Device\\Management} &319 & 2549 & 826 &564	&1128	&266\\\cline{2-9}
    &\textbf{04} & \makecell[c]{Network Service\\Management}&4 & 36 & 25 &16	&32	&13\\\cline{2-9}
    &\textbf{05} & \makecell[c]{Intelligent Optical\\Network} &5701 & 31173 & 6145 &14543	&29086	&2649 \\\cline{2-9}
    &\textbf{06} & \makecell[c]{Automated\\ O \& M}&235 & 235 & 223 &1506	&3012	&373\\\cline{2-9}
    &\textbf{07} & \makecell[c]{Optical Virtual\\Private Network}&131 & 131 & 124 &584	&1168	&337\\
    \hline
    \textbf{Total} & & & \textbf{6984} & \textbf{37816} & \textbf{9963} &\textbf{17825}& \textbf{35818} &\textbf{4277}\\
    \hline
    \end{tabular}
    }
    \vspace{-0.5cm} 
    \label{table: datasets}
\end{table*}
We construct a comprehensive dataset composed of 13 public datasets and 7 proprietary industrial datasets provided by Huawei. The public datasets originate from diverse software domains---including healthcare (iTrust\cite{gao2024triad}), governance (eANCI\cite{hey2021improving}), supply chain management (viper\cite{ferrari2017pure}), education (SMOS\cite{hey2021improving}), and others---exhibiting high syntactic complexity, such as frequent use of subordinate clauses, nested conditions, and intricate sentence structures.
The requirements in industrial datasets, written in Simplified Chinese with simple and concise sentence structures, are collected from the Network Cloud Engine-Transport (NCE-T) product. This product has been deployed over the past decade in network and service management, disaster recovery, automated operations and maintenance, and optical networking.
\cref{table: datasets} summarizes the key statistics of these datasets.
In principle, the number of alternative flows should correspond to the number of branch points---one for each alternative flow.
However, among 9 of the 13 public datasets that contain alternative flows, only one provides branch point annotations. Therefore, the implicit branch points in the other 8 datasets need to be annotated.

\subsubsection{Branch Point Annotation}
\begin{figure}[]
    \centering
    \includegraphics[width=1.0\linewidth]{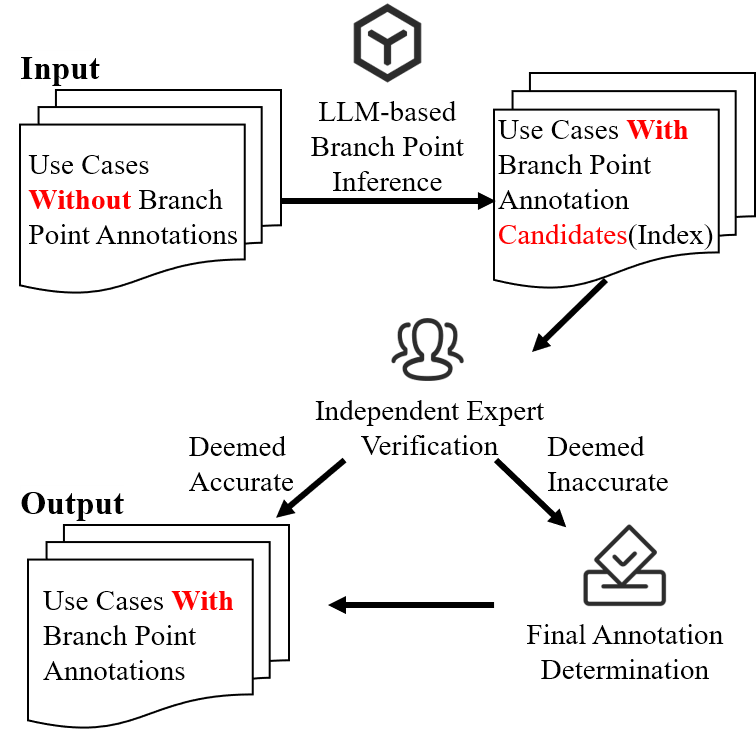}
    \setlength{\abovecaptionskip}{0.cm} 
    \caption{Semi-automated Branch Point Annotation}
    \vspace*{-0.5cm} 
    \label{fig. Branch Point Completion}
\end{figure}

Accurate generation of alternative flows relies on precise branch point annotations.
To annotate missing branch points in 8 public datasets, we design a semi-automated pipeline that integrates LLM-based inference with expert validation. 
For every use case that contains alternative flows without explicit branch point steps, we recover the branch points using the procedure illustrated in \cref{fig. Branch Point Completion}.

\textbf{LLM-based Branch Point Inference.}
Inputs are the use case description, the base flow, and the corresponding alternative flow. The LLM is prompted to identify and output the most plausible step index in the base flow where the corresponding alternative flow diverges, i.e., its branch point.

\textbf{Independent Expert Verification.}
The LLM-inferred branch point suggestions are independently reviewed by five software engineers, with industrial experience in requirements analysis and development ranging from six to ten years.
They evaluate if the predicted branch points accurately correspond to the branching semantics indicated by the alternative flow. 
If an engineer deems the LLM's prediction inaccurate, they are required to propose the appropriate branch point(s). 

\textbf{Final Annotation Determination.}
For each alternative flow, the LLM-predicted branch point is retained if approved by a majority of engineers. Otherwise, the final branch point is determined through consensus among the engineers’ candidate proposals.

\textbf{Annotation Reliability Analysis.}
To assess the quality and reliability of the branch point annotations, we report the inter-annotator agreement (the degree of consensus among annotators) and LLM–human alignment on the annotations. 
First, we compute Fleiss’ $\kappa$ \cite{fleiss1971measuring} over all annotated branch points to quantify the level of agreement among the five experts. This provides a standard measure of annotation consistency. 
Second, we measure the alignment between the engineers’ final annotations and the LLM’s predictions.
Notably, a Fleiss’ $\kappa$ value of 1 indicates identical annotations from all five experts, while an LLM–Human alignment value of 1 indicates that the LLM's prediction exactly matches the experts' final agreed branch points.

\cref{table: Dataset Statistics} summarizes the number of alternative flows without branch points in each dataset (i.e., the number of branch points to be annotated), the corresponding Fleiss’ $\kappa$ among the five expert annotators, and the proportion of LLM-suggested branch points matching the final results. 
As shown in \cref{table: Dataset Statistics}, the average Fleiss’ $\kappa$ of 0.778 indicates substantial inter-annotator agreement, validating the reliability of manual annotations. The average LLM–human alignment score of 0.888 demonstrates the effectiveness of LLM-based branch point prediction.
Together, these results demonstrate that the proposed semi-automated annotation pipeline produces highly accurate branch point annotations. The final branch-point annotations and the corresponding extended dataset files are released as part of the replication package.

\begin{table}[htbp]
\centering
\setlength{\abovecaptionskip}{0.cm} 
\caption{Annotation Reliability and LLM Alignment for Branch Points.}
\scalebox{0.9}{  
\begin{tabular}{lccc}
\hline
\textbf{Dataset}       & \multicolumn{1}{c}{\textbf{\begin{tabular}[c]{@{}c@{}}\#AF\\ Missing BP\end{tabular}}} & \multicolumn{1}{c}{\textbf{Fleiss’ $\kappa$}} & \multicolumn{1}{c}{\textbf{\begin{tabular}[c]{@{}c@{}}LLM–Human\\ Alignment\end{tabular}}} \\ \hline
\textbf{keepass}       & 37                                                                                     & 0.677                                      & 0.865                                                                                      \\
\textbf{gamma j}       & 14                                                                                     & 1.000                                          & 0.929                                                                                      \\
\textbf{inventory}     & 12                                                                                     & 0.567                                      & 0.917                                                                                      \\
\textbf{hats}          & 1                                                                                      & 1.000                                          & 1.000                                                                                          \\
\textbf{pnnl}          & 9                                                                                      & 1.000                                          & 1.000                                                                                          \\
\textbf{viper}         & 18                                                                                     & 0.691                                      & 0.889                                                                                      \\
\textbf{inventory 2.0} & 21                                                                                     & 0.561                                      & 0.905                                                                                      \\
\textbf{iTrust}        & 5                                                                                      & 0.724                                      & 0.6                                                                                        \\ \hline
\textbf{Overall(Mean)}         & \textbf{117}                                                                           & \textbf{0.778}                             & \textbf{0.888}                                                                             \\ \hline
\end{tabular}
}
\vspace{-0.5cm} 
\label{table: Dataset Statistics}
\end{table}

\begin{table*}[ht]
\centering
\setlength{\abovecaptionskip}{0.cm} 
\caption{Node Statistics for Public and Industrial Datasets}
\scalebox{0.9}{  
\begin{tabular}{llcccccc}
\hline
                                    & \multicolumn{1}{c}{\multirow{2}{*}{\textbf{Tool}}} & \multicolumn{2}{c}{\textbf{Basic Flow}}                               & \multicolumn{2}{c}{\textbf{Alternative Flow}}                        & \multicolumn{1}{c}{\textbf{Use Case Description}} & \multicolumn{1}{c}{\multirow{2}{*}{\textbf{Sum}}} \\ \cline{3-4} \cline{5-7} 
                                    & \multicolumn{1}{c}{}                               & \multicolumn{1}{l}{\# of Actions} & \multicolumn{1}{l}{\# of Objects} & \multicolumn{1}{l}{\# of Actions} & \multicolumn{1}{l}{\# of Object} & \multicolumn{1}{c}{\# of Core words}                                  & \multicolumn{1}{l}{}                     \\ \hline
\textbf{\multirow{3}{*}{\makecell{Public\\Datasets}}}    & Stanford CoreNLP                                   & 684                               & 1277                              & 217                               & 181                              & 1123                                                  & \textbf{3482}                                     \\
                                    & ERNIE 4.0 Turbo                                    & 908                               & 1387                              & 182                               & 156                              & 1261                                                  & \textbf{3894}                                     \\
                                    & GPT-4o                                             & 786                               & 1464                              & 190                               & 268                              & 1311                                                  & \textbf{4019}                                     \\
\hline
\textbf{\multirow{3}{*}{\makecell{Industrial\\Datasets}}} & Stanford CoreNLP                                   & 851                               & 1070                              & 349                               & 331                              & 1305                                                  & \textbf{3906}                                     \\
                                    & ERNIE 4.0 Turbo                                    & 778                               & 6429                              & 179                               & 1943                             & 5265                                                  & \textbf{14594}                                    \\
                                    & GPT-4o                                             & 710                               & 4535                              & 293                               & 921                              & 5144                                                  & \textbf{11603}                
\\\hline                  
\end{tabular}
}
\vspace{-0.5cm} 
    \label{table: node Statistics}
\end{table*}

\subsubsection{Dataset Statistics}
Since \methodALL relies on the extracted core words, actions, and objects to construct semantic relations and supervise the enhanced R-GAT model, its performance is highly sensitive to the richness of those elements present in the datasets. 
As shown in \cref{table: node Statistics}, we compare the number of nodes extracted by the mature NLP toolkit, Stanford CoreNLP, and two representative LLMs, ERNIE 4.0 Turbo and GPT-4o.
The numbers of extracted nodes from industrial datasets (3,906 / 14,594 / 11,603) are larger than those from public datasets (3,482 / 3,894 / 4,019). This indicates that industrial datasets may contain richer domain-specific content, such as specialized terms, data objects, and actions.
Moreover, the two LLMs can extract a greater variety and quantity of elements than the traditional NLP tool---14,594 and 11,603 compared to 3,906 in the industrial datasets---justifying the motivation for employing LLMs in \methodALL.
This richer set of extracted elements further enhances the construction of the SRG, particularly the representation effectiveness of the R-GAT encoder.
The number of edges (i.e., relations between nodes) is not reported separately, since the edge count is inherently dependent on the number of nodes. Specifically, the counts for $E_1$--$E_2$, $E_4$--$E_8$ are directly related to the number of nodes extracted, while the count for $E3$ is derived from the number of action steps in a use case flow.

\subsection{Evaluation Metrics} \label{Evaluation Metrics}
To quantitatively evaluate the effectiveness of \methodALL, we employ multiple metrics to evaluate its three modules---\methodBFGen, \methodBPPredictor, and \methodAFGen separately.

\subsubsection{Metrics for \methodBFGen and \methodAFGen}
\methodBFGen and \methodAFGen generate use case flows from given requirement descriptions. 
We employ the four well-known metrics: $Precision$, $Recall$, $F1$ $score$, and $Area$ $Under$ $Curve$ ($AUC$) to evaluate the effectiveness and performance of the two modules, based on the matches between the generated use case flows and the corresponding reference flows. 

\begin{equation}
\begin{aligned}
    \text{$Precision$} = \frac{|\text{$R$} \cap \text{$G$}|}{|\text{$G$}|}, \text{$Recall$} = \frac{| \text{$R$} \cap \text{$G$} |}{|\text{$R$}|},\\
    \text{$F1$} = \frac{2 \times \text{$Precision$} \times \text{$Recall$}}{\text{$Precision$} + \text{$Recall$}}
    \label{eq:precision}
\end{aligned}
\end{equation}
where $G$ denotes the set of nodes generated by \methodBFGen or \methodAFGen, and $R$ denotes the set of nodes in the corresponding reference flow.
$Precision$ measures the accuracy, $Recall$ measures the completeness, and $F1$ measures the overall balance of the generation.

Literal requirement descriptions are the common input of \methodBFGen and \methodAFGen. Due to the inherent ambiguity of natural language, similar words might have very different meanings in different domain contexts. To quantitatively assess how well the modules differentiate near-synonymous terms and detect domain term mismatches in different domains, we introduce $AUC$ as an evaluation metric, calculated as follows:
\begin{equation}
\begin{split}
\text{$AUC$} = \frac{1}{|P|\cdot|N|} \sum_{i \in P} \sum_{j \in N} \Bigl[ & \mathbb{I}\big(p(i) > p(j)\big) \\
& + 0.5 \times \mathbb{I}\big(p(i) = p(j)\big) \Bigr]
\end{split}
\label{eq:AUC}
\end{equation}
where \(P\) and \(N\) denote positive (i.e., matched in the reference flow) and negative (i.e., not matched) nodes, respectively. \(p(i)\) is the predicted probability that node \(i\) belongs to the generated flow. \(\mathbb{I}(\cdot)\) is the indicator function. If node \(i\) is present in the outputs of \methodBFGen or \methodAFGen, \(p(i)\) will be 1.0, otherwise 0.

\subsubsection{Metrics for Branch Point Prediction (\methodBPPredictor)}\label{metric:bpp}
Branch points might occur in any step of a base flow. Normally, branch points are sparsely and unevenly distributed in use case flows. A use case flow may have zero to the maximal number of action steps as branch points. A single global metric is insufficient for evaluating the effectiveness of branch point prediction. Therefore, we employ $micro$- averaged $Precision$, $Recall$, and $F1$ $score$ to assess overall step-level prediction performance, and $macro$- averaged $Precision$, $Recall$, and $F1$ $score$ to ensure fair evaluation across individual base flows, particularly those that are short or contain few branch points.
For each use case $uc \in UC$:
\begin{equation}
\begin{aligned}
    \text {$Precision$}_{uc}=\frac{\left|BP_{uc}^{\text {pred}} \cap BP_{uc}^{\text {ref }}\right|}{\left|BP_{uc}^{\text {pred}}\right|}, \\
    \text{$Recall$}_{uc}=\frac{\left|B P_{uc}^{\text {pred}} \cap BP_{uc}^{\text {ref }}\right|}{\left|BP_{uc}^{\text {ref}}\right|}, \\
    F 1_{ \text {$uc$}}=\frac{2 \times \text {$Precision$}_{\text {$uc$}} \times \text {$Recall$}_{\text {$uc$}}}{\text {$Precision$}_{\text {$uc$}}+\text {$Recall$}_{ \text {$uc$}}}
\end{aligned}
\end{equation}
where $BP_{uc}^{\text {pred}}$ and $BP_{uc}^{\text {ref}}$ denote the predicted and reference branch point set for use case $uc$ respectively.
The macro-averaged and micro-averaged metrics are computed as follows, where $\lvert UC \rvert$ is the number of use cases.
\begin{equation}
\begin{aligned}
    \text {$Precision$}_{ \text {$macro$}}=\frac{1}{|UC|} \sum_{uc \in UC} \text {$Precision$}_{uc}, \\
    \text {$Recall$}_{ \text {$macro$}}=\frac{1}{|UC|} \sum_{uc \in UC} \text {$Recall$}_{uc},\\
    F 1_{ \text {$macro$}}=\frac{1}{|UC|} \sum_{uc \in UC} \text {$F1$}_{uc}
\end{aligned}
\end{equation}
\begin{equation}
\begin{aligned}
    \text {$Precision$}_{ \text {$micro$}}=\frac{\sum_{uc \in UC}\left|BP_{uc}^{\text {pred }} \cap BP_{uc}^{\text {ref }}\right|}{\sum_{uc \in UC}\left|B P_{uc}^{\text {pred }}\right|}, \\ 
    \text {$Recall$}_{ \text {$micro$}}=\frac{\sum_{uc \in UC}\left|BP_{uc}^{\text {pred }} \cap BP_{uc}^{\text {ref }}\right|}{\sum_{uc \in UC}\left|BP_{uc}^{\text {ref }}\right|}, \\
    F 1_{ \text {$micro$}}=\frac{2 \times \text {$Precision$}_{\text {$micro$}} \times \text {$Recall$}_{\text {$micro$}}}{\text {$Precision$}_{\text {$micro$}}+\text {$Recall$}_{ \text {$micro$}}}
\end{aligned}
\end{equation}

\subsection{Research Questions (RQs) \& Baseline Setup} \label{sec:rq set}

\textbf{\textit{RQ1:}} How effective is \methodBFGen in generating basic flows compared to the baseline methods?

The goal of RQ1 is to investigate whether the basic flows generated by \methodBFGen are of high quality and whether it has advantages over the baselines.
To answer this question, we design a comparative experiment to compare the performance of \methodBFGen against the four baseline methods that lie in three categories: rule-based method, LLM-based method, and GNN-based method. We select three representative rule-based methods \cite{jahan2021generating,20163402722828,yue2015atoucan}. 
Given that some rules in these methods are tailored for specific input formats, we consolidate the broadly applicable rules into a unified framework. 
Additionally, we replace the earlier NLP tools used in these methods for parsing requirement descriptions with the more mature Stanford CoreNLP\cite{manning2014stanford} to feed those methods with consistent inputs.
LLM-based methods are currently considered the most prominent ones. Since our datasets are specified in English and Simplified Chinese, we select two established models---GPT-4o from OpenAI \cite{key} and ERNIE 4.0 Turbo from Baidu \cite{baiduERNIE40Turbo128KModelBuilder}---to minimize the impact of linguistic differences on the LLMs' performance.
Since \methodBFGen uses the R-GAT model, we select the GNN-based RGAT-with-BERT method \cite{meng2023rgat}, which is related to \methodBFGen and has shown strong performance, as a baseline. This method utilizes pre-trained BERT to generate hybrid representations and employs the R-GAT model to encode syntactic dependency graphs and learn syntactic embeddings.

\textbf{\textit{RQ2:}} How effective is the SIP module in contributing to the overall performance of \methodBFGen?

To ensure the enhanced R-GAT can precisely model the various nodes and relations, we design the SIP module to extract the semantic information in terms of core words, actions, and objects, and their relations. RQ2 aims to evaluate the validity of our approach by investigating whether the SIP module contributes positively to the overall effectiveness of \methodBFGen.

We design an ablation study to evaluate the impact of SIP on \methodBFGen, and set \methodBFGen without SIP (\methodBFGen w/o SIP) as the first baseline for comparison with the full \methodBFGen. 
We further include a second baseline, \methodBFGen without preprocessing (\methodBFGen w/o Preproc.), to evaluate the contribution of the two preprocessing tasks: Sentence Simplification and Splitting.
It should be noted that we select Stanford CoreNLP as the text parsing tool for providing semantic information in \methodBFGen w/o SIP. 
Since \methodBFGen w/o Preproc. relies on LLMs to extract semantic information, we evaluate it with two distinct models---GPT-4o and ERNIE 4.0 Turbo---to mitigate the potential randomness by LLM.

\textbf{\textit{RQ3:}} How effective are different values of the attention preservation factor for \methodBFGen?

As specified in \cref{section:bfgen}, we introduce the attention preservation factor, $\lambda$, to enable the R-GAT to capture important relations. RQ3 systematically examines how $\lambda$ affects the efficacy of \methodBFGen.
We conduct a hyperparameter sensitivity experiment to examine the impact of $\lambda$ on the effectiveness of \methodBFGen, evaluating \methodBFGen's performance as $\lambda$ varies over its considered range.

\textbf{\textit{RQ4:}} How does the completeness of requirements affect the effectiveness of \methodBFGen?

In modern software engineering practices, especially in agile software development, requirements are progressively elaborated through iterations \cite{russo2021agile}. At the beginning, requirements briefly specify what a system should do, without details on processing flow, data definition and validation. This type of requirement is usually called a high-level requirement \cite{heimdahl2002completeness}. 
Since \methodBFGen aims to generate the basic flow from a given use case description, the completeness of the use case description may affect its effectiveness. Therefore, it is necessary to evaluate \methodBFGen with various levels of incompleteness in the use case description.

We design a requirement completeness sensitivity experiment to evaluate \methodBFGen's performance with incomplete requirements by implementing Random Masking \cite{huang2022self} of content words in requirements to simulate different levels of completeness. By increasing the number of masked content words, we simulate requirements with varying degrees of incompleteness, using fully complete requirements (100\%) as the baseline.

\textbf{\textit{RQ5:}} How effective is the \methodBPPredictor in identifying branch points in base flows?

To assess how effectively \methodBPPredictor identifies branch points in given base flows, we conduct a comparative experiment against a set of representative baselines. To the best of our knowledge, there is no published research directly focused on predicting branch points in use case flows. The most relevant studies, based on our literature review, are rule-based event identification techniques that recognize conditional steps in requirement descriptions \cite{jurkiewicz2015automated, williams2022automated, rago2016assisting, makino2008method}. However, these approaches are tightly coupled with specific datasets that are not publicly accessible. 
Moreover, these studies do not provide the necessary implementation code or the specific rules they employ, making replication or adaptation to the datasets in this research impossible. 
Therefore, we consider two categories of baselines: LLM-based baselines and a Structure-Agnostic Sequence Transformer baseline. 
For the LLM-based baselines, we use GPT-4o and ERNIE 4.0 Turbo.
GPT-4o is a highly competitive model with strong overall performance, while ERNIE 4.0 Turbo addresses potential shortcomings of GPT-4o in processing Simplified Chinese text.
To further strengthen the comparison, we implement a Structure-Agnostic Sequence Transformer (Sequence Transformer) baseline. This baseline uses the same node features, supervision signals, data splits, and evaluation protocol as our approach, but replaces SRG-based graph propagation with sequence modeling over the ordered base flow nodes. It encodes the ordered base flow node sequence with a Transformer encoder and predicts branch-point labels for each base flow node via a feed-forward classification head.

\textbf{\textit{RQ6:}} How effective is \methodAFGen in generating alternative flows given a use case description, its base flow and the corresponding branch points?

To address RQ6, we design a comparative experiment to evaluate \methodAFGen against baseline methods.
To the best of our knowledge, the only method we found for alternative flow generation (Ko et al. \cite{20163402722828}), which is rule-based, was published 10 years ago. The necessary dataset, rules, and implementation details are no longer accessible, which prevents replication.
Given this lack of reproducibility, we use GPT-4o, ERNIE 4.0 Turbo, and the previously introduced Structure-Agnostic Sequence Transformer (Sequence Transformer) baseline for comparison. For alternative flow generation, the Sequence Transformer baseline employs a Transformer decoder conditioned on the branch point and its fixed 1-hop local context to autoregressively generate the alternative flow action sequence.

\textbf{\textit{RQ7:}} How does the scope of contextual information used to initialize the decoder’s hidden state affect the quality of generated alternative flows?

The initialization of the decoder’s hidden state plays a key role in the quality of alternative flows generated by \methodAFGen. 
Specifically, it encodes contextual information from the $k$-hop neighborhood of the branch point, where the parameter $k$ governs the scope of contextual aggregation and thus directly affects \methodAFGen’s ability to capture both local and long-range dependencies in the SRG.

To investigate RQ7, we conduct a hyperparameter sensitivity experiment on the context scope parameter $k$ with respect to the effectiveness of \methodAFGen.
Specifically, we evaluate \methodAFGen under five values of $k$, ranging from node-local ($k=0$) to graph-global of $k=all$:
(1) 0-hop ($k=0$): aggregates only the branch point’s own embedding, without any contextual information;
(2) 1-hop ($k=1$): uses information from directly connected neighbors;
(3) 2-hop ($k=2$): uses information from second-order neighbors;
(4) 3-hop ($k=3$): uses information from third-order neighbors;
(5) All nodes ($k=all$): uses all information in the graph.
We evaluate the impact of each $k$ setting on the quality of generated alternative flows using standard metrics: $F1$ $score$ and $AUC$.

\subsection{Experiment Settings} \label{sec: experiment environment}
To conduct the designed experiments and to answer the research questions, we perform systematic grid searches to identify the optimal parameter configurations that maximize task-specific performance. 

In the experiments for answering RQ1 and RQ3, the learning rate for \methodBFGen was set to 0.1, and the dropout rate was set to 0.3. In the experiments for answering RQ2, the learning rate was set to 1e-4, and the dropout rate was set to 0.2. In the experiments for answering RQ4, the learning rate was set to 0.01, and the dropout rate was set to 0.35. 

For the experiments answering RQ5 and RQ6, we use different settings based on the dataset. Specifically, for the industrial NCE-T datasets, the learning rate was set to 2e-4, and the attention preservation factor $\lambda$ was set to 0.8, while for public datasets, the learning rate was set to 1e-4, and $\lambda$ to 0.84. To prevent overfitting, the early stopping patience was fixed at 20 epochs across all experimental runs.

For all the experiments except those investigating the sensitivity of the parameter $\lambda$ and the specific configurations for RQ5 and RQ6 mentioned above, the value of $\lambda$ was fixed at 0.9. Threshold $\tau$ is set to 0.85. The threshold $\theta$ for determining whether a node is positive is set to 0.5. Regarding the usage of LLMs in baselines and the SIP module, the temperature parameter was set to the default value (1.0) to ensure result stability.

The experiments were conducted on a server equipped with 22 vCPUs (Intel(R) Xeon(R) Platinum 8470Q), 110GB of RAM, and a single NVIDIA RTX PRO 6000 GPU (96GB), running on a Linux operating system.

For reproducibility, all raw experimental data, branch point annotations, baseline implementations, and other materials in the experiments are included in the replication package, available at https://github.com/WGYbuaa/FlowGen.

\section{Results and Analysis} \label{sec: results}
\subsection{RQ1: How effective is \methodBFGen in generating basic flows compared to the baseline methods?}

\begin{table*}[htbp]
\small 
    \setlength{\abovecaptionskip}{0cm} 
	\setlength{\belowcaptionskip}{-0.2cm} 
    
    \centering
    \caption{Comparative Experimental Results of \methodBFGen and Baselines on Basic Flow Generation:\\$Precision$, $Recall$, $F1$ $Score$ and $AUC$ Across Public and Industrial Datasets - RQ1}
    \scalebox{0.9}{  
    \begin{tabular}{llcccc}
        \hline
        \textbf{Dataset} & \textbf{Approach} & \textbf{Precision} & \textbf{Recall} & \textbf{F1} & \textbf{AUC} \\
        \hline
        \multirow{5}{*}{\textbf{Public Datasets}} & Rule-based & 0.364 & 0.180 & 0.215 & 0.090 \\
        & ERNIE 4.0 Turbo & 0.510 & 0.407 & 0.417 & 0.204 \\
        & GPT-4o & 0.371 & 0.288 & 0.279 & 0.144 \\
        & R-GAT with BERT & 0.229 & 0.118 & 0.156 & 0.708 \\
        & \methodBFGen & 0.582 / \textbf{+14.12\%} & 0.508 / \textbf{+24.82\%} & 0.543 / \textbf{+30.22\%} & 0.846 / \textbf{+19.49\%} \\
        \hline
        \multirow{5}{*}{\textbf{Industrial Datasets}} & Rule-based & 0.200 & 0.242 & 0.202 & 0.121 \\
        & ERNIE 4.0 Turbo & 0.347 & 0.209 & 0.240 & 0.104 \\
        & GPT-4o & 0.221 & 0.156 & 0.166 & 0.078 \\
        & R-GAT with BERT & 0.543 & 0.454 & 0.494 & 0.786 \\
        & \methodBFGen & 0.621 / \textbf{+14.36\%} & 0.488 / \textbf{+7.49\%} & 0.547 / \textbf{+10.73\%} & 0.865 / \textbf{+10.05\%} \\
        \hline
    \end{tabular}
    }
    \vspace{-0.5cm} 
    \label{tab:dataset_comparison}
\end{table*}

\cref{tab:dataset_comparison} presents the comparative experimental results between \methodBFGen and the four baselines. These results represent the average values obtained from several independent experiments, indicating that, among all methods, \methodBFGen achieves the best performance on both public and industrial datasets.

As shown in \cref{tab:dataset_comparison}, the rule-based baseline method exhibits limited effectiveness. 
Despite using mature NLP toolkits like Stanford CoreNLP, the method fails to handle the linguistic ambiguity and differentiate terms with domain-specific meanings. Moreover, some pre-defined rules are too rigid to adapt to different presentation styles and domain scenarios.
In contrast, \methodBFGen leverages the SIP module---enabled by LLMs with robust text processing capabilities and rich domain knowledge---to achieve significant improvements across all four metrics on both public and industrial datasets. 

The performance of LLM baseline methods, ERNIE 4.0 Turbo and GPT-4o, both with robust generation capabilities, yields similar results, slightly surpassing the rule-based methods. 
However, after rigorously analyzing the basic flows generated by the two LLMs, we identified hallucinations. Some action steps in the generated basic flows operate out of system boundaries. This violates the fundamental restriction of the use case.
For instance, in the eANCI dataset, a use case describes the system's functionality of presenting knowledge about fire causes to the public. However, the LLM generates a basic flow describing firefighters' fire suppression procedures and the operational mechanisms of fire protection systems, which are relevant to firefighters' work but not to this case, as they are not dictated in the given functional requirement descriptions.
This indicates that the two LLMs fail to adequately incorporate the necessary contextual information from the functional descriptions. This is the key reason why \methodBFGen outperforms LLM methods.
The SIP module equipped with an LLM in \methodBFGen focuses on extracting terms, actions, and objects from requirement descriptions. They are then fed into the enhanced R-GAT to capture the necessary contextual information. The LLM does not directly participate in step sequence generation, effectively mitigating the hallucination inherent in LLMs.

The large scale and high domain knowledge density of the industrial datasets provide rich information for GNN training, enabling both the R-GAT with BERT baseline and \methodBFGen to achieve better performance.
However, the R-GAT with BERT baseline cannot adjust the importance of node connections according to domain and requirement context---a limitation that hinders further performance improvement.
To overcome the pitfalls, \methodBFGen obtains more precise information with the SIP module using LLM, and proposes the attention preservation factor to dynamically adjust the influence of edge weights to prioritize the most contextually relevant node connections, thereby further achieving enhanced accuracy of capturing semantic nuances and domain-specific matches.

\subsection{RQ2: How effective is the SIP module in contributing to the overall performance of \methodBFGen?}

\begin{figure}
    \centering
\setlength{\abovecaptionskip}{0.cm} 
    \includegraphics[width=\linewidth]{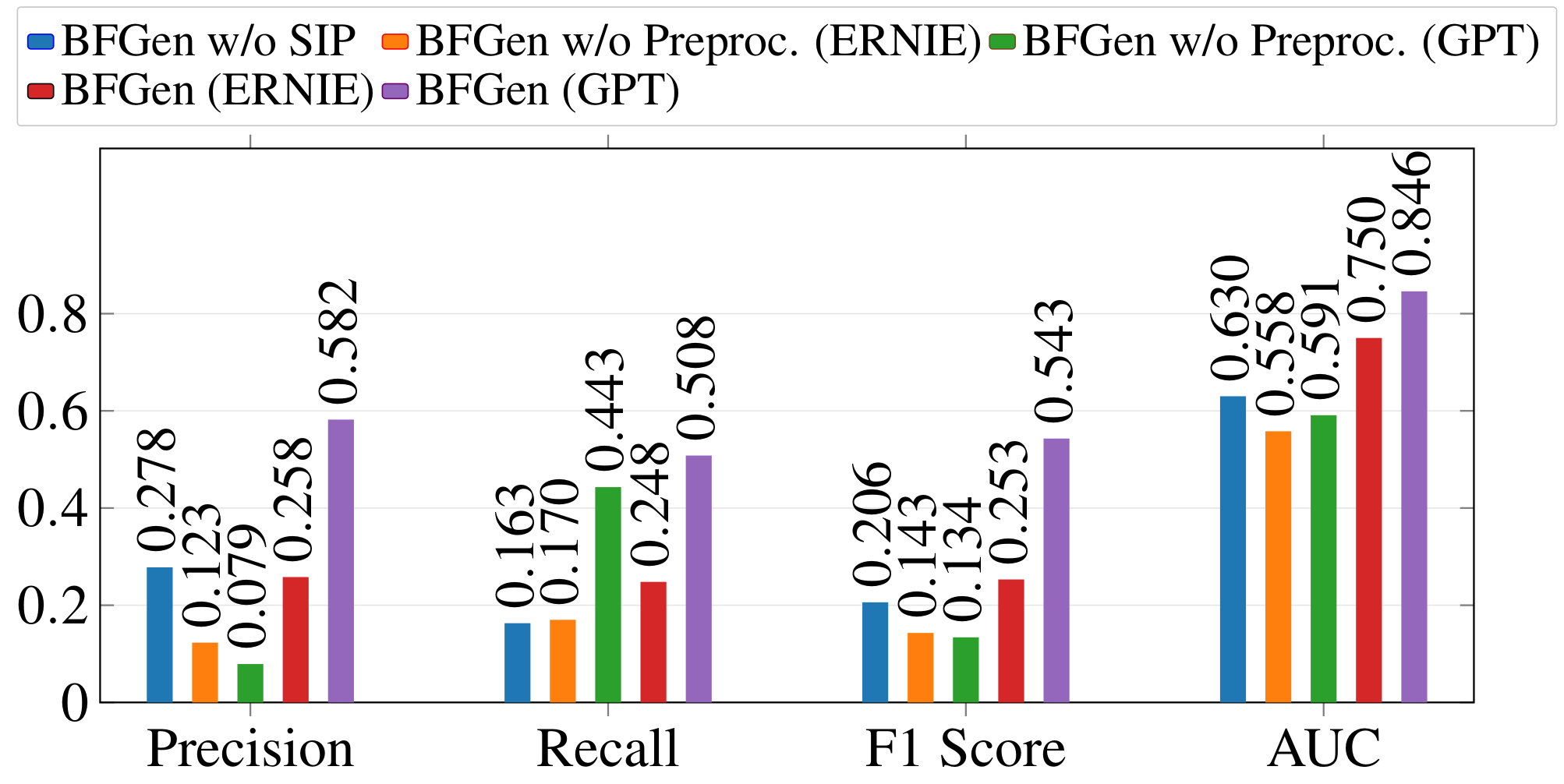}
    \setlength{\abovecaptionskip}{0.cm} 
    \caption{RESULTS OF ABLATION STUDY FOR SIP MODULE - RQ2. Note: Preproc. = Preprocessing.}
    \label{RQ2 p1}
    \vspace{-0.5cm} 
\end{figure}

We conduct this ablation study on public datasets, which exhibit higher syntactic complexity than industrial datasets and thus provide a more stringent test for validating the effectiveness of the SIP module. As shown in \cref{RQ2 p1}, both variants of the complete \methodBFGen ---equipped with LLMs---achieve strong performance, with \methodBFGen (GPT) outperforming all other methods.


\methodBFGen outperforms \methodBFGen w/o SIP. Compared to Stanford CoreNLP, the rich domain knowledge and extensive training of LLMs integrated in the SIP allow \methodBFGen to handle domain-specific terms and matches, greatly enhancing the accuracy of information extraction, which is essential for the subsequent modeling and training of the enhanced R-GAT.


\methodBFGen also outperforms \methodBFGen w/o Preproc.
Though \methodBFGen w/o Preproc. performs well in extracting semantic elements from simple sentences or sentences with few modifiers, its performance sharply decreases when dealing with complex sentences and compound sentences.
Some complex sentences containing multiple verbs and nouns are difficult to parse for definitive action-object relations. 
In contrast, the complete \methodBFGen with sentence simplification and splitting can systematically parse and disambiguate input sentences (e.g., clarifying verbs and their accessed objects), greatly reducing the noise in the data and the complexity of information extraction.

The experimental results demonstrate that the LLM-equipped SIP module and the two preprocessing tasks, Sentence Simplification and Splitting, contribute significantly to enhancing the performance of \methodBFGen.
Interestingly, we observe that certain LLM-enabled baseline methods (e.g., \methodBFGen w/o Preproc. (ERNIE)) underperform \methodBFGen without SIP on certain metrics such as F1 score. This is primarily due to their unstable output and intrinsic hallucinations that lead to excessive extraction of inaccurate information in complex sentences, which induces erroneous or missing relations, ultimately degrading the overall performance.

\subsection{RQ3: How effective are different values of the attention preservation factor for \methodBFGen?}

\begin{figure}
    \centering
\setlength{\abovecaptionskip}{0.cm} 
    \includegraphics[width=\linewidth]{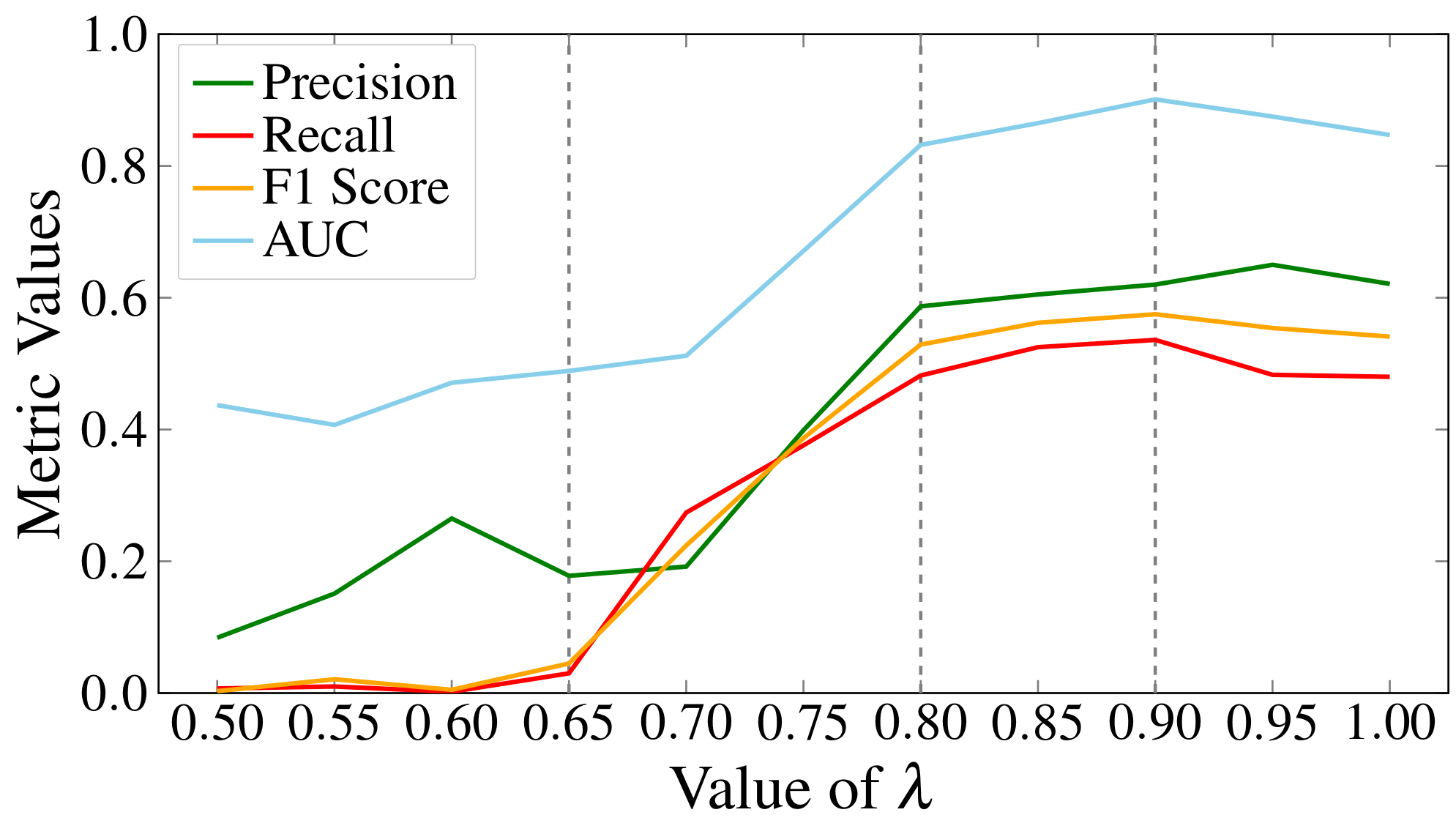}
    \setlength{\abovecaptionskip}{0.cm} 
    \caption{Performance vs. Attention Preservation Factor ($\lambda$)- RQ3}
    \label{RQ3 p1}
    \vspace{-0.5cm} 
\end{figure}

We conduct hyperparameter sensitivity experiments on industrial datasets to investigate the impact of the attention preservation factor $\lambda$ on \methodBFGen.
Compared to public datasets, industrial datasets contain more domain-specific matches---such as sequential actions and fixed interactions---which render model performance more sensitive to variations in $\lambda$.
Notably, when $\lambda$ \textless 0.5, the model exhibits a sharp decline in both convergence stability and generalization capability.
Consequently, our analysis focuses on the performance behavior for $\lambda$ $\geq$ 0.5.

\cref{RQ3 p1} illustrates that \methodBFGen’s performance is highly sensitive to $\lambda$, and the resulting curve can be segmented into four phases:
(1) When $\lambda$ is in the range of [0.50, 0.65), lower values of $\lambda$ weaken the contextual information capture and modeling ability of the R-GAT module, leading to low performance. (2) As $\lambda$ increases to [0.65, 0.80), the model can better distinguish the importance of various connections in the data with the captured contextual information, resulting in a significant performance improvement. (3) When $\lambda$ is in the range of [0.80, 0.90), the model performance reaches a high level since the differences between different relationships captured in the model are used in adequate learning and training. (4) When $\lambda$ is set in the range of [0.90, 1], model performance decreases. 
This stems from overly focusing on local details and small variations, which amplifies sensitivity to noise in the data.
Therefore, for datasets with high domain knowledge density, the model achieves a balance between attention-based and uniform embedding aggregation when $\lambda$ $\in$ [0.80, 0.90), effectively emphasizing critical relations in the context while mitigating the impact of noise from distant information.

\subsection{RQ4: How does the completeness of requirement affect the effectiveness of \methodBFGen?}

\begin{figure}
    \centering
    \setlength{\abovecaptionskip}{0.cm} 
    \includegraphics[width=\linewidth]{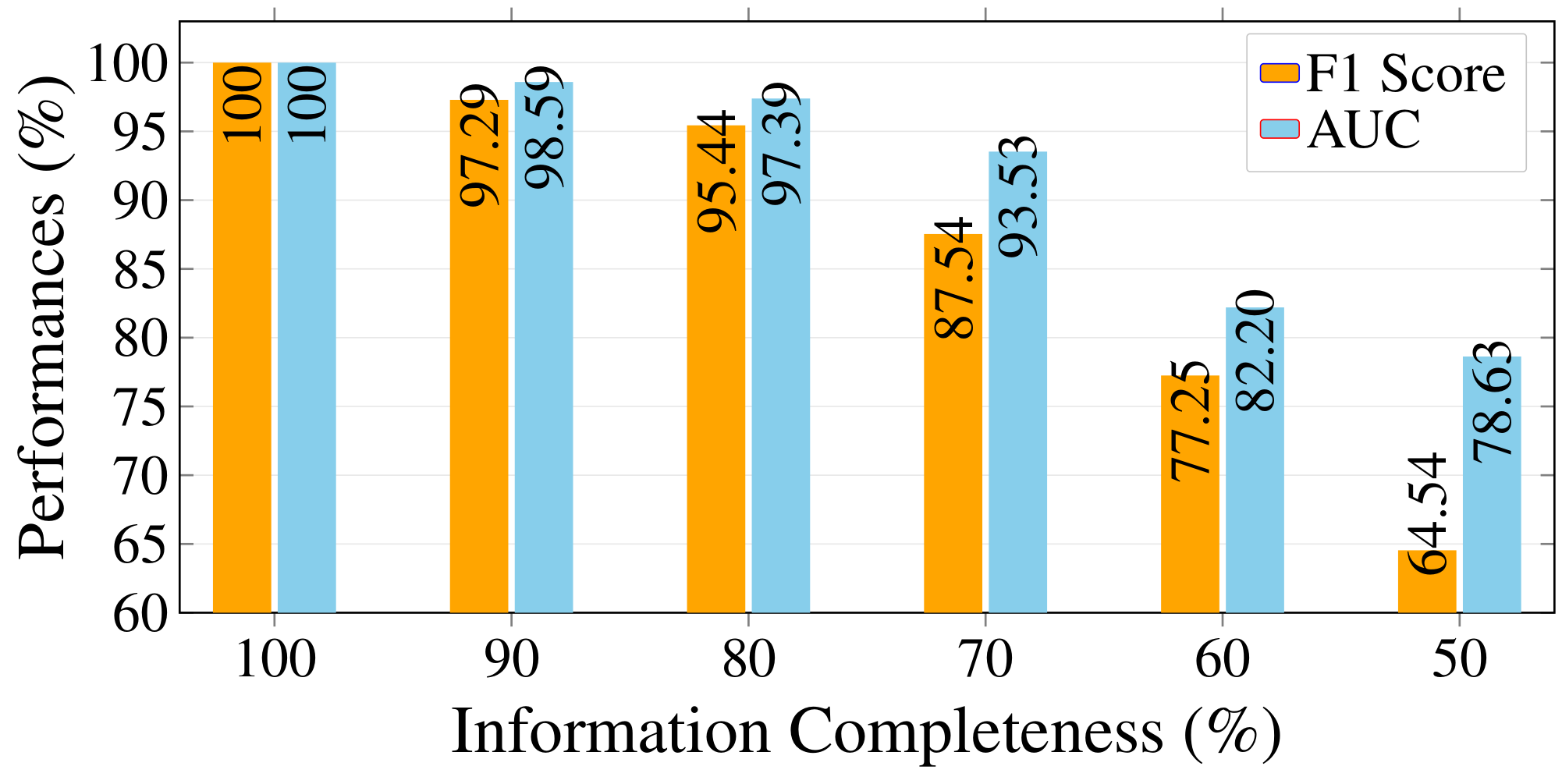}
    \setlength{\abovecaptionskip}{0.cm} 
    \caption{Results of Requirement Completeness Sensitivity Experiments - RQ4}
    \label{RQ4 p1}
    \vspace{-0.5cm} 
\end{figure}

\cref{RQ4 p1} presents the results of sensitivity analysis on requirement completeness. The five datasets on the right show the performance retention of \methodBFGen (relative to the baseline, in percentage) as requirement completeness decreases. The findings indicate that:
(1) With at least 80\% of the information retained, \methodBFGen's performance shows only slight fluctuation (F1 and AUC drop less than 5\%), indicating strong tolerance to insufficient information. This provides strong evidence of its applicability in practical industrial scenarios since it is hard to ensure the completeness of requirements; 
(2) As the completeness decreases further, \methodBFGen gradually loses its ability to recognize certain key actions or objects, particularly in the scenarios with the absence of domain-specific or contextual core words; 
(3) The slower decline rate of AUC compared to F1 score as the completeness decreases indicates that \methodBFGen consistently assigns higher probabilities to positive samples, suggesting its underlying feature ranking capability remains stable. 
Therefore, the experimental results demonstrate that \methodBFGen exhibits strong robustness to the completeness of requirement descriptions, maintaining stable performance even when completeness drops to 80\%.

\subsection{RQ5: How effective is the \methodBPPredictor in identifying branch points in base flows?}

\begin{table*}[htbp]
\centering
\setlength{\abovecaptionskip}{0.cm} 
\caption{Comparative Experimental Results of \methodBPPredictor and Baselines On Branch Point Prediction
:\\$Precision$, $Recall$ and $F1$ $Score$ Across Public and Industrial Datasets - RQ5.}
\scalebox{0.8}{ 
\begin{tabular}{llcccccc}
\hline
\multirow{2}{*}{\textbf{Dataset}}                                              & \multirow{2}{*}{\textbf{Approach}} & \multicolumn{3}{c}{\textbf{Macro-Average}}                 & \multicolumn{3}{c}{\textbf{Micro-Average}}                  \\ \cline{3-5}  \cline{6-8}
                                                                               &                                    & \textbf{Precision} & \textbf{Recall}   & \textbf{F1 Score} & \textbf{Precision} & \textbf{Recall}    & \textbf{F1 Score} \\ \hline
\multirow{5}{*}{\begin{tabular}[c]{@{}l@{}}Public\\ Datasets\end{tabular}}     & GPT-4o                             & 0.333              & 0.278             & 0.300             & 0.097              & 0.214              & 0.133             \\
                                                                               & ERNIE 4.0 Turbo                    & 0.400              & 0.600             & 0.467             & 0.184              & 0.500              & 0.269             \\
                                                                               & Sequence Transformer              & 0.180              & 0.383             & 0.244             &
                                                                               0.238              & 0.417           & 0.303                \\ 
                                                                               & \methodBPPredictor \textit{(GPT)}                          & 0.389              & 0.794             & 0.501             & 0.402              & 0.726              & 0.518             \\
                                                                               & \methodBPPredictor \textit{(ERNIE)}                        & \textbf{0.657/+64.25\%}   & \textbf{0.900/+50.00\%} & \textbf{0.717/+53.53\%}  & \textbf{0.500/+110.08\%}  & \textbf{0.956/+91.20\%}     & \textbf{0.657/+116.83\%}   \\ \hline
\multirow{5}{*}{\begin{tabular}[c]{@{}l@{}}Industrial\\ Datasets\end{tabular}} & GPT-4o                             & 0.198              & 0.074             & 0.097             & 0.220              & 0.022              & 0.040             \\
                                                                               & ERNIE 4.0 Turbo                    & 0.376              & 0.206             & 0.230             & 0.302              & 0.076              & 0.122             \\
                                                                               & Sequence Transformer              & 0.550              & 0.732             & 0.599             &
                                                                               0.600              & 0.738           & 0.662                \\ 
                                                                               & \methodBPPredictor \textit{(GPT)}                          & 0.747              & 0.937             & 0.818             & 0.740              & 0.973              & 0.840             \\
                                                                               & \methodBPPredictor \textit{(ERNIE)}                        & \textbf{0.792/+44.00\%}  & \textbf{0.976/+33.33\%} & \textbf{0.869/+45.08\%} & \textbf{0.782/+30.33\%}  & \textbf{0.986/+33.60\%} & \textbf{0.873/+31.87\%} \\ \hline

\end{tabular}
}
\label{table:rq5}
\end{table*}

As shown in \cref{table:rq5}, \methodBPPredictor significantly outperforms all baseline methods across all metrics on both datasets, demonstrating consistently robust performance. This clearly highlights its dual strengths: effectively handling a wide range of use cases (including those with dense or sparse branch points), while maintaining high overall prediction accuracy. 

Further analysis reveals an interesting contrast: the two LLM-based baselines perform better on public datasets than on industrial datasets, whereas both the Sequence Transformer baseline and \methodBPPredictor exhibit the opposite trend, achieving superior performance on the industrial datasets. We attribute this divergence to the intrinsic characteristics of the two datasets. 
The industrial datasets contain many more branch points with similar syntactic structure, more regular basic flow patterns, and substantially more training samples, enabling supervised models to learn stable local and global regularities more effectively. In particular, the Sequence Transformer baseline benefits from these data characteristics because branch points in the industrial datasets are often associated with recurring local sequential cues that can be captured by sequence modeling alone. In contrast, the public datasets are more flexible and diverse in syntax, domain, and branching style, and the relatively limited training data constrains the modeling capacity of both \methodBPPredictor and the Sequence Transformer baseline.
Meanwhile, the LLM baselines perform worse on the industrial datasets primarily because it exhibits much more domain-specific knowledge and contains more specialized terms, while general-purpose LLMs lack the necessary prior training for such industrial scenarios, making accurate prediction challenging.

Further analysis of the LLM-based baselines' outputs shows that a small number of errors stem from step-index misalignment.
For example, LLMs generated the content of step 2 but labeled it as step 3---even though step indices were explicitly annotated in the use case flow and the prompt explicitly instructed them to use zero-based indexing for all predictions, which is crucial for use case flows with repeated operations.
These errors were introduced due to LLMs’ inherent limitations as sequence-to-sequence models: they lack an explicit mechanism to align input steps with output labels \cite{liu2024llms}, revealing a mismatch between the linear, context-only modeling paradigm of LLMs and the structured, position-sensitive nature of our task.
In contrast, \methodBPPredictor leverages an enhanced R-GAT encoder to model use cases as a graph rather than text sequences, capturing both semantics and structural dependencies, and thereby inherently avoiding such misalignment errors.

Compared with the Sequence Transformer baseline, \methodBPPredictor also maintains a distinct advantage, suggesting that branch-point identification depends not only on local sequential cues but also on branch-triggering control-flow logic and its relation to the surrounding flow context, both of which are explicitly propagated over the SRG rather than left implicit in sequence-only modeling.

\subsection{RQ6: How effective is \methodAFGen in generating alternative flows given a use case description, its base flow and the corresponding branch points?}

\begin{table*}[h]
\centering
\setlength{\abovecaptionskip}{0.cm} 
\caption{Comparative Experimental Results of \methodAFGen and Baselines on Alternative Flow Generation
:\\$Precision$, $Recall$, $F1$ $Score$ and $AUC$ Across Public and Industrial Datasets - RQ6.}
\scalebox{0.9}{  
\begin{tabular}{llcccc}
\hline
\textbf{Dataset}                                                                        & \textbf{Approach} & \textbf{Precision}         & \textbf{Recall}          & \textbf{F1 Score}                & \textbf{AUC}               \\ \hline
\multirow{5}{*}{\textbf{\begin{tabular}[c]{@{}l@{}}Public\\ Datasets\end{tabular}}}     & GPT-4o            & 0.195                      & 0.568                    & 0.244                      & 0.284                      \\
                                                                                        & ERNIE 4.0 Turbo   & 0.165                      & \textbf{0.680}           & 0.245                      & 0.340                      \\
                                                                                        & Sequence Transformer & 0.241                  & 0.416                    & 0.305
                                                                                                 &0.826                       \\                                                      
                                                                                        & \methodAFGen \textit{(GPT)}       & 0.286                      & 0.272                    & 0.279                      & \textbf{0.831 / +0.61\%} \\
                                                                                        & \methodAFGen \textit{(ERNIE)}     & \textbf{0.297 / +23.24\%}  & 0.455 / \textbf{-33.09\%} & \textbf{0.359 / +17.70\%}  & 0.827                      \\ \hline
\multirow{5}{*}{\textbf{\begin{tabular}[c]{@{}l@{}}Industrial\\ Datasets\end{tabular}}} & GPT-4o            & 0.121                      & \textbf{0.431}           & 0.176                      & 0.216                      \\
                                                                                        & ERNIE 4.0 Turbo   & 0.063                      & 0.410                    & 0.104                      & 0.205                      \\
                                                                                        & Sequence Transformer & 0.712                  & 0.403                    & 0.515
                                                                                                 & 0.930                       \\ 
                                                                                        & \methodAFGen \textit{(GPT)}       & 0.584                      & 0.430 / \textbf{-0.23\%} & 0.495                      & 0.941                      \\
                                                                                        & \methodAFGen \textit{(ERNIE)}     & \textbf{0.772 / +8.43\%} & 0.419                    & \textbf{0.543 / +5.44\%} & \textbf{0.953 / +2.47\%} \\ \hline
\end{tabular}
}
\vspace{-0.3cm} 
\label{table:rq6}
\end{table*}

As shown in \cref{table:rq6}, \methodAFGen significantly outperforms the baseline models in Precision, F1 score, and AUC on both datasets, demonstrating its effectiveness in generating high-quality alternative flows. However, it achieves slightly lower Recall compared to the LLM-based baselines.
Further analysis of the LLM baselines’ outputs reveals that they tend to produce overly verbose alternative flows containing redundant or irrelevant steps. 
Taking the public datasets as an example, statistics show that in the test set, the alternative flows generated by the GPT-4o baseline have an average of 11 action steps, while those generated by ERNIE 4.0 Turbo have an average of 15.79 steps. In contrast, the ground truth alternative flows contain only an average of 4.79 steps.
This so-called "over-generation" strategy increases the number of matches with actions and objects, thereby inflating recall. However, it introduces a large number of incorrect predictions, severely compromising precision, as reflected in the substantially lower Precision and F1 scores.

In contrast, \methodAFGen generates more accurate alternative flows by leveraging the enhanced R-GAT to extract neighboring node information around branch points, thus producing representations that capture both semantic and structural dependencies. The generated alternative flows are not only semantically sound and consistent with the original use case logic, but also achieve a balance between coverage and accuracy.

Additionally, the experimental results for RQ6 exhibit a trend similar to that observed in RQ5: the LLM-based baselines perform better on public datasets than on industrial datasets, whereas the Sequence Transformer baseline and \methodAFGen achieve superior performance on industrial datasets. As discussed in RQ5, industrial datasets contain a large number of structurally regular alternative flows, with partial similarity in their content, and provide abundant training samples, enabling supervised models to learn more effectively. 
Both the Sequence Transformer baseline and \methodAFGen benefit from these data characteristics. The Sequence Transformer baseline can already model a considerable portion of these flows from local sequential patterns alone, whereas \methodAFGen further benefits from graph-based modeling of flow dependencies, which makes it more robust to interference from domain-specific terms in the textual descriptions.
In contrast, the limited training data in public datasets restricts \methodAFGen’s ability to generalize from diverse alternative operations, while the LLM-based baselines, benefiting from pretraining on large-scale general corpora, demonstrate stronger generalization capabilities, leading to relatively better performance on these datasets.

The Sequence Transformer baseline is a stronger comparator than the LLM baselines and is competitive with \methodAFGen. Nevertheless, \methodAFGen still maintains an overall advantage. This can be attributed in part to SRG-based graph conditioning, which better preserves branch-to-flow alignment and requirement logic—including semantic consistency, control-flow logic, and data-flow logic—than sequence-only modeling.

\subsection{RQ7: How does the scope of contextual information used to initialize the decoder’s hidden state affect the quality of generated alternative flows?}

\begin{figure}
    \centering
    \setlength{\abovecaptionskip}{0.cm} 
    \includegraphics[width=\linewidth]{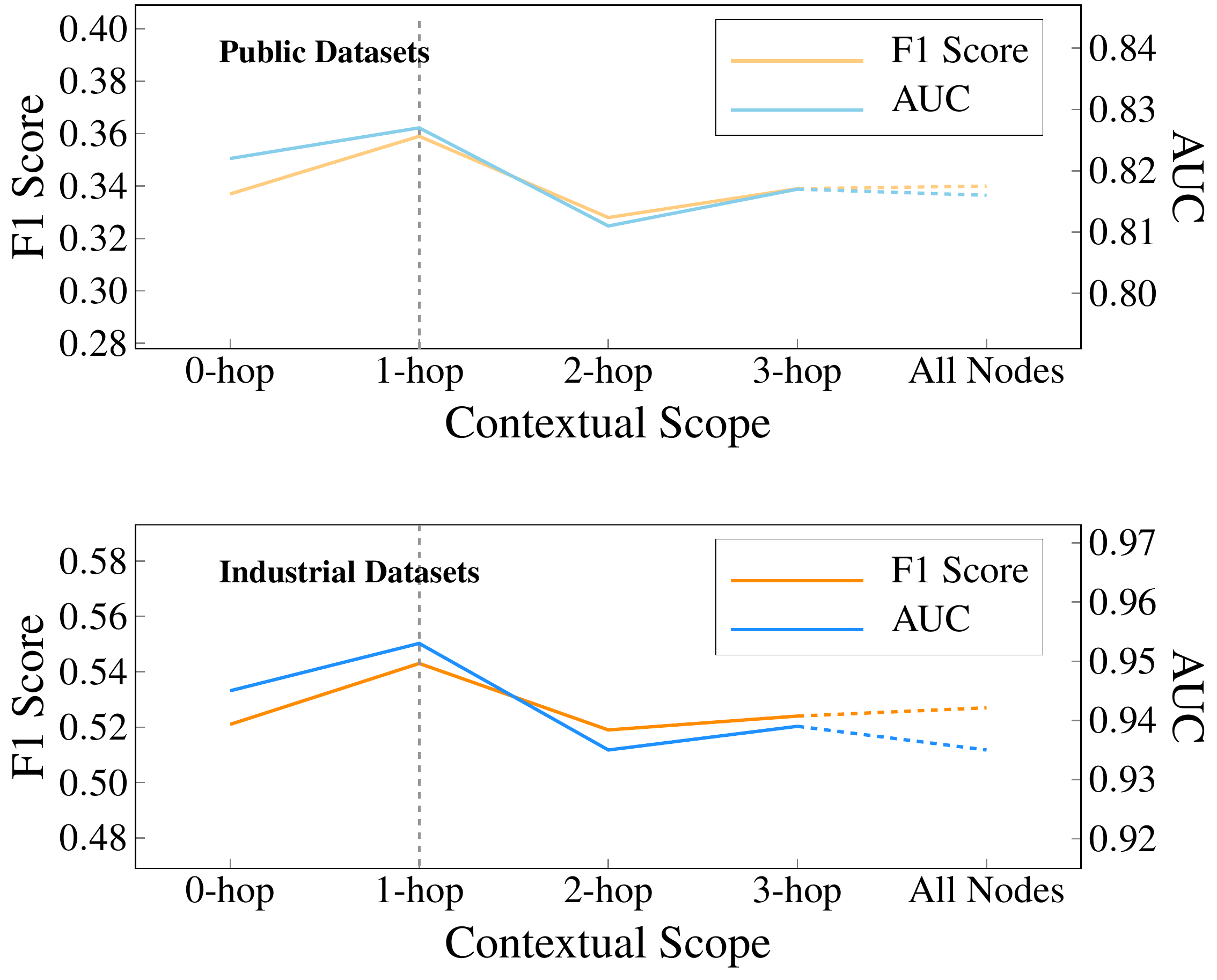}
    \caption{Performance vs. Contextual Scope ($k$) - RQ7.}
    \label{pic: RQ7}
    \vspace{-0.5cm} 
\end{figure}

As shown in \cref{pic: RQ7}, the performance of \methodAFGen (ERNIE) exhibits a clear trend when using different scopes of context information: both F1 score and AUC peak at the 1-hop neighborhood, while dropping significantly when extending to the context of 2-hop, 3-hop, or All Nodes. 
This suggests that incorporating close neighbors' information is most effective for guiding alternative flow generation.
The poor performance of the 0-hop setting---which uses only the branch point’s embedding without any contextual input---confirms the importance of local structural and semantic cues in capturing functional dependencies within the SRG. 
In contrast, when the context range extends beyond 1-hop, redundant or noisy information from distant nodes may degrade model performance. The All Nodes setting, in particular, underperforms the 1-hop setting, suggesting that aggregating over all nodes may blur the fine-grained semantic distinctions critical for accurate step prediction.
These results imply that a moderate scope of context---specifically, 1-hop neighbors---is optimal for initializing the decoder in \methodAFGen. It strikes a balance between leveraging sufficient local semantics and avoiding interference from irrelevant long-range dependencies.


\section{Threats To Validity}\label{sec: Threats to Validity}
Threats to the internal validity pertain to experimental biases and errors that may originate from four primary sources: language unification of public datasets, natural language parsing tools, the branch-point annotation process, and the hyperparameter configurations set during model training. 
(1) Given the presence of Italian and English in public datasets, we follow the methodology proposed by Hey et al. \cite{hey2021improving}, utilizing the high-performance translation engine DeepL \cite{deeplDeepLTranslate} to unify the language into English. This process effectively mitigates potential cross-lingual bias risks. We integrate the translated dataset into the replication package to ensure methodological transparency.
(2) To mitigate potential biases stemming from limited NLP accuracy, we leverage excellent LLMs, GPT-4o and ERNIE 4.0 Turbo, to enhance the SIP module's capabilities. However, we acknowledge the inherent probabilistic nature of generative models. Even with fixed hyperparameters (e.g., temperature), minor variations in output may occur. To minimize this threat, we conducted multiple runs for critical steps and reported averaged results.
(3) In the branch point annotation pipeline, the five engineers review the LLM suggestions independently of one another, but the review is not blind to the suggested branch point. This may introduce anchoring bias. Nevertheless, the substantial Fleiss' $\kappa$ and the high LLM-human alignment suggest that the final annotations remain reasonably stable.
(4) During the modeling and training, critical hyperparameters---including $\lambda$, early stopping epochs, and learning rates---as well as the initialization context scope for \methodAFGen’s hidden state have a substantial impact on model performance. The hyperparameter settings in our experiments were derived from results of the grid searches we conducted. However, we explicitly acknowledge that hyperparameter optimization requires more comprehensive empirical validation. In subsequent research phases, we plan to conduct dedicated empirical studies to establish evidence-based guidelines for hyperparameter selection.

Threats to external validity primarily concern the generalizability of \methodALL. For this study, we rigorously validated the proposed approach using 13 public datasets and 7 industrial datasets spanning diverse domains. This multi-source validation strategy surpasses conventional academic benchmarks confined to singular domains, thereby better approximating real-world engineering scenarios. 
Notably, these datasets collectively contain 6,984 use cases, with the largest sub-dataset containing 5,701 use cases—already exhibiting characteristics of a large-scale dataset.
While current empirical validation demonstrates \methodALL's preliminary generalizability, adhering to the "more-is-better" principle, we plan to implement \methodALL on more diverse datasets across additional domains.

Threats to construct validity stem from the appropriateness of evaluation metrics. 
We employ Precision, Recall, F1 score, and AUC as the assessment framework. Precision, Recall, and F1 score are representative multi-objective classification metrics \cite{zhang2013review}, widely adopted as performance indicators for models \cite{gao2024triad}. AUC is included as a complementary indicator of discriminative performance \cite{8246530}. 
Moreover, to address the class imbalance inherent in the branch point prediction task, we report results using both macro- and micro-average metrics.
These metrics provide a fair and consistent basis for comparing \methodALL with all baseline methods.
Nevertheless, they mainly assess overlap and discrimination at the node or branch-point level, and do not fully capture higher-level properties of complete use case flows, such as end-to-end executability.

\section{Conclusion} \label{sec: conclusion}

This paper presents \methodALL for complete use case flow generation from high-level requirements. \methodALL integrates LLM-based semantic extraction, SRG-based relational modeling, basic flow generation, branch point prediction, and alternative flow generation, together with a semi-automatic pipeline for supplementing branch-point annotations in public datasets.

Evaluations on 13 public and 7 industrial datasets demonstrate that \methodALL outperforms baseline methods across major metrics. In particular, for branch point prediction and alternative flow generation, the comparisons against LLM-based baselines and the Sequence Transformer baseline indicate that SRG-based graph conditioning is useful for preserving branch-to-flow alignment and requirement logic beyond sequence-only modeling.
Ablation studies and hyperparameter sensitivity experiments validate the effectiveness of the LLM-based semantic information processing module and the introduced attention preservation factor in \methodBFGen, while also revealing the impact of contextual scope on \methodAFGen’s performance. Furthermore, a requirement completeness sensitivity experiment confirms that \methodBFGen maintains robust performance even when input requirements are incomplete.

Future work will focus on exploring how preconditions and postconditions in use case specifications can be effectively modeled and used to guide the generation of both basic and alternative flows, and how more explicit system boundary representations can be introduced to further constrain out-of-scope generation.

\section*{Acknowledgments}
This work was supported in part by the Integration and Application of Digital Learning Technology Ministry of Education Innovation under Grant 1431005.

\printcredits

\bibliographystyle{model1-num-names}

\bibliography{ref}

\end{document}